\documentclass[twocolumn]{aastex631}

\usepackage{amsmath}

\begin{document}

\title{JWST-MIRI Spectroscopy of Warm Molecular Emission and Variability in the AS 209 Disk}

\correspondingauthor{Carlos E. Romero-Mirza}
\email{ carlos.romero\_mirza@cfa.harvard.edu [Change]}

\author[0000-0001-7152-9794]{Carlos E. Romero-Mirza}
\affiliation{Center for Astrophysics $\vert$ Harvard \& Smithsonian, Cambridge, MA 02138, USA}

\author[0000-0001-8798-1347]{Karin I. Öberg}
\affiliation{Center for Astrophysics $\vert$ Harvard \& Smithsonian, Cambridge, MA 02138, USA}

\author[0000-0003-4335-0900]{Andrea Banzatti}
\affiliation{Department of Physics, Texas State University, 749 N Comanche Street, San Marcos, TX 78666, USA}

\author[0000-0001-7552-1562]{Klaus M. Pontoppidan}
\affiliation{Space Telescope Science Institute, 3700 San Martin Dr., Baltimore, MD, 21218, USA}
\affiliation{Jet Propulsion Laboratory, California Institute of Technology, 4800 Oak Grove Dr. Pasadena, CA, 91109, USA}

\author[0000-0003-2253-2270]{Sean M. Andrews}
\affiliation{Center for Astrophysics $\vert$ Harvard \& Smithsonian, Cambridge, MA 02138, USA}

\author[0000-0003-1526-7587]{David J. Wilner}
\affiliation{Center for Astrophysics $\vert$ Harvard \& Smithsonian, Cambridge, MA 02138, USA}

\author[0000-0003-4179-6394]{Edwin A. Bergin}
\affiliation{Department of Astronomy, University of Michigan, 323 West Hall, 1085 S. University Avenue, Ann Arbor, MI 48109, USA}

\author[0000-0002-1483-8811]{Ian Czekala}
\affiliation{Department of Astronomy and Astrophysics, The Pennsylvania State University, University Park, PA 16802, USA}
\affiliation{Center for Exoplanets and Habitable Worlds, The Pennsylvania State University, University Park, PA 16802, USA}
\affiliation{Center for Astrostatistics, 525 Davey Laboratory, The Pennsylvania State University, University Park, PA 16802, USA}
\affiliation{Institute for Computational \& Data Sciences, The Pennsylvania State University, University Park, PA 16802, USA}

\author[0000-0003-1413-1776]{Charles J. Law}
\affiliation{Center for Astrophysics $\vert$ Harvard \& Smithsonian, Cambridge, MA 02138, USA}
\affiliation{Department of Astronomy, University of Virginia, Charlottesville, VA 22904, USA}
\altaffiliation{NASA Hubble Fellowship Program Sagan Fellow}

\author[0000-0003-3682-6632]{Colette Salyk}
\affiliation{Vassar College, 124 Raymond Avenue, Poughkeepsie, NY 12604, USA}

\author[0000-0003-1534-5186]{Richard Teague}
\affiliation{Department of Earth, Atmospheric, and Planetary Sciences, Massachusetts Institute of Technology, Cambridge, MA 02139, USA}

\author[0000-0001-8642-1786]{Chunhua Qi}
\affiliation{Center for Astrophysics $\vert$ Harvard \& Smithsonian, Cambridge, MA 02138, USA}

\author[0000-0002-8716-0482]{Jennifer B. Bergner}
\affiliation{University of Chicago Department of the Geophysical Sciences, Chicago, IL 60637, USA}

\author[0000-0001-6947-6072]{Jane Huang}
\affiliation{Department of Astronomy, University of Michigan, 323 West Hall, 1085 S. University Avenue, Ann Arbor, MI 48109, USA}
\affiliation{Department of Astronomy, Columbia University, 538 W. 120th Street, Pupin Hall, New York, NY, United States of America}

\author[0000-0001-6078-786X]{Catherine Walsh}
\affiliation{School of Physics and Astronomy, University of Leeds, Leeds, LS2 9JT, UK}

\author[0000-0003-4784-3040]{Viviana V. Guzmán}
\affiliation{Instituto de Astrofísica, Pontificia Universidad Católica de Chile, 7820436 Macul, Santiago, Chile}
\affiliation{N\'ucleo Milenio de Formación Planetaria (NPF), Gran Bretaña 1111, Valparaiso, Chile}

\author[0000-0003-2076-8001]{L. Ilsedore Cleeves}
\affiliation{Astronomy Department, University of Virginia, Charlottesville, VA 22904, USA}

\author[0000-0003-3283-6884]{Yuri Aikawa}
\affiliation{Department of Astronomy, Graduate School of Science, The University of Tokyo, Tokyo 113-0033, Japan}

\author[0000-0001-7258-770X]{Jaehan Bae}
\affiliation{Department of Astronomy, University of Florida, Gainesville, FL 32611, USA}

\author[0000-0003-2014-2121]{Alice S. Booth}
\affiliation{Leiden Observatory, Leiden University, 2300 RA Leiden, The Netherlands}

\author[0000-0002-2700-9676]{Gianni Cataldi}
\affiliation{National Astronomical Observatory of Japan, 2-21-1 Osawa, Mitaka, Tokyo 181-8588, Japan }

\author[0000-0003-1008-1142]{John D. Ilee}
\affiliation{School of Physics and Astronomy, University of Leeds, Leeds, LS2 9JT, UK}

\author[0000-0003-1837-3772]{Romane Le Gal}
\affiliation{Université Grenoble Alpes, CNRS, IPAG, F-38000 Grenoble, France}

\author[0000-0002-7607-719X]{Feng Long}
\affiliation{Center for Astrophysics $\vert$ Harvard \& Smithsonian, Cambridge, MA 02138, USA}
\affiliation{Lunar and Planetary Laboratory, University of Arizona, Tucson, AZ 85721, USA}
\altaffiliation{NASA Hubble Fellowship Program Sagan Fellow}

\author[0000-0002-8932-1219]{Ryan A. Loomis}
\affiliation{National Radio Astronomy Observatory, 520 Edgemont Road, Charlottesville, VA 22903, USA}

\author[0000-0002-1637-7393]{François Menard}
\affiliation{Université Grenoble Alpes, CNRS, IPAG, F-38000 Grenoble, France}

\author[0000-0002-7616-666X]{Yao Liu}
\affiliation{Purple Mountain Observatory \& Key Laboratory of Radio Astronomy, Chinese Academy of Sciences, Nanjing 210033, China}

\begin{abstract}

We present MIRI MRS observations of the large, multi-gapped protoplanetary disk around the T-Tauri star AS 209. The observations reveal hundreds of water vapor lines from 4.9 to 25.5 $\mu$m towards the inner $\sim1$ au in the disk, including the first detection of ro-vibrational water emission in this disk. The spectrum is dominated by hot ($\sim800$ K) water vapor and OH gas, with only marginal detections of CO$_2$, HCN, and a possible colder water vapor component. Using slab models with a detailed treatment of opacities and line overlap, we retrieve the column density, emitting area, and excitation temperature of water vapor and OH, and provide upper limits for the observable mass of other molecules. Compared to MIRI spectra of other T-Tauri disks, the inner disk of AS 209 does not appear to be atypically depleted in CO$_2$ nor HCN. Based on \textit{Spitzer IRS} observations, we further find evidence for molecular emission variability over a 10-year baseline. Water, OH, and CO$_2$ line luminosities have decreased by factors 2-4 in the new MIRI epoch, yet there are minimal continuum emission variations. The origin of this variability is yet to be understood.

\end{abstract}



\section{Introduction} \label{sec:intro}

The presence of water in circumstellar disks plays a fundamental role in the outcome of planet formation. Water is expected to be a major oxygen carrier in disks---therefore key to set the local gas-phase and solid C/O and N/O ratios in different disk regions \citep[e.g.][]{Herbst_2009, van_Dishoeck_2014, Oberg_2021}---and undeniably an essential ingredient for setting planetary habitability. In the outer disk, beyond the water snowline, the stickiness of water ice (among other factors) is thought to facilitate the growth of pebbles and planetesimals \citep{Dominik_1997, Birnstiel_2010, Gundlach_2015}. Moreover, the migration efficiency of these bodies might be regulated by the local concentration of water ice \citep{Stevenson_1988, Ciesla_2006}. This solid water reservoir, however, is not observable in most disks. Instead, water can be most readily characterized via its gas-phase emission spectrum, which arises from the warm, innermost disk regions (within 10 au) in the mid-IR \citep{Pontoppidan_2014}, and from the outer disk at far-IR wavelengths \citep{Hogerheijde_2011, Du_2017}. 

Our current view of inner-disk water comes primarily from the Infrared Spectrograph onboard the Spitzer Space Telescope (\textit{Spitzer IRS}) \citep{Houck_2004}, which revealed that bright, mid-IR rotational water emission is ubiquitous among low-mass T Tauri stars \citep{Carr_2008, Pontoppidan_2010}. In general, the intensity of mid-IR water emission lines in these sources is best reproduced by column densities of order $10^{18}$ cm$^{-2}$, warm excitation temperatures ranging from $T_{ex} \approx 200$ to $800$ K, and emitting areas in the disk of around 10 au$^2$, corresponding to outer circular radii of $\lesssim 2$ au \citep{Carr_2011, Salyk_2011, Najita_2011, Antonellini_2015}. These emitting areas tend to be consistent with line kinematic information inferred from high-resolution ground-based spectra \citep{Pontoppidan_2010b, Banzatti_2014, Najita_2018, Salyk_2019, Banzatti_2023}. Mid-IR observations have further revealed that the intensity of water emission is anti-correlated to the mass and evolutionary stage of the central star \citep{Pontoppidan_2010, Salyk_2011, Fedele_2011}. Determining the abundance and distribution of water vapor in disks can thus provide substantial insights regarding the disk's dynamical and thermal evolution. 

The characterization of water vapor and the trends described above have so far encountered two critical constraints. First, due to the low spectral resolution and sensitivity of \textit{Spitzer IRS}, most of the water lines observed with this instrument are blended, and model fits were dominated by the brightest, most optically thick water lines that emit near the disk surface \citep[e.g.][]{Salyk_2011, Carr_2011}. The end result is that slab models infer $10-100\times$ lower abundances relative to the total water column predicted by thermochemical simulations that account for chemical heating and H$_2$O self-shielding \citep{Meijerink_2009, Blevins_2016, Woitke_2019, Bosman_2022}. 

Second, a complete understanding of water vapor in disks has been precluded by the limited wavelength coverage of \textit{Spitzer IRS} \citep{Banzatti_2023}. While \textit{Spitzer IRS} could readily detect bright rotational lines tracing lukewarm (400-600 K) water vapor in disks, a hotter ($T_{ex} \sim 1000$ K), more compact water reservoir is only revealed by ro-vibrational water emission within 2-8 $\mu$m \citep{Carr_2004, Salyk_2008, Mandell_2012}, which had been only accessible to ground-based spectrographs. Similarly, a cold water reservoir ($T_{ex} \approx 200$ K) associated with sublimation near the snowline and photochemistry in the outer regions of some disks was mainly observable with far-IR facilities like \textit{Herschel} \citep{Zhang_2013, Du_2017}. No single instrument until now has enabled a comprehensive analysis of distinct water emission reservoirs with sufficient sensitivity and spectral resolution. 

Thanks to its unparalleled capabilities, the James Webb Space Telescope (JWST) has the potential to overcome the difficulties mentioned above and accurately characterize excitation conditions of water vapor in disks. In particular, the Mid-Infrared Instrument \citep[MIRI;][]{Rieke_2015} Medium-resolution Spectrometer \citep[MRS;][]{Wells_2015} offers medium spectral resolution ($R = 1500-3700$) from 4.9 to 28 $\mu$m, allowing the simultaneous analysis of ro-vibrational (bending mode) and rotational water lines spanning a vast range of upper energy levels ($E_u = 10^3 - 10^4$ K) for the first time. Similarly, the sensitivity of MIRI provides the opportunity to identify and characterize multiple optically thin water lines, and hence trace water closer to the planet-forming disk midplane. Indeed, JWST MIRI has already started to provide robust evidence for a low-energy water line excess in compact disks \citep{Banzatti_2023b} as well as the first detections of weak isotopologue lines \citep{Grant_2022}, small organics \citep{kospal_2023}, and ionization tracers \citep{berne_2023} in the terrestrial-planet forming region of disks.

In this paper, we model the emission of H$_2$O, OH radical (hereafter OH), CO$_2$, C$_2$H$_2$, and HCN---with particular focus on water vapor and OH---using MIRI MRS data of the inner disk of AS~209. The disk of AS~209 is the first of four targets observed as part of the JWST follow-up to the ALMA MAPS Large Program \citep{Oberg_2021b}, and the only one with prior inner-disk chemistry constraints from \textit{Spitzer IRS} in this sample. In Section \ref{sec:obs}, we describe the observations and data reduction process. The spectral modeling technique and retrieval framework are detailed in Section \ref{sec:analysis}, and Section \ref{sec:results} presents the excitation analysis results. In Section \ref{sec:discussion} we explore whether the mid-IR spectrum of AS 209 differs from other T-Tauri disks, and compare our findings with prior \textit{Spitzer IRS} observations. We find evidence for significant variability in molecular emission features, and present some possible explanations for this anomaly. 

\section{Observations} \label{sec:obs}

\subsection{The Disk of AS 209}

The low-mass \citep[$0.96 M_{\odot}$;][]{Fang_2018} T Tauri star AS~209, located at 121 pc \citep{Gaia}, hosts an unusually bright and extended protoplanetary disk, with a mm-dust radius of $R_{dust} = 139$ au and a CO $2-1$ gas disk that extends out to 272 au \citep{Andrews_2018, Huang_2018, guzman_2018}. Based on ALMA observations, the disk exhibits a least seven deep dust gaps from 9 out to 137 au \citep{Huang_2018, Sierra_2021}. Many molecules have been detected in the AS 209 outer disk (beyond 10 au) at sub-millimiter wavelengths \citep[e.g.][]{oberg_2011, Huang_2017, Bergner_2018, Oberg_2021}, with most exhibiting multiple ring-like emission substructures when observed at high spatial resolution \citep{Law_2021}. 

There is evidence that at least one of the dust gaps is associated with ongoing planet formation at 100 au \citep{Favre_2019}, and a candidate circumplanetary disk has been detected in molecular gas at $\sim 200$ au \citep{Bae_2022}. The possibility that additional planets may be forming within the innermost $\sim 10$ au makes AS~209 an attractive target for IR follow-up observations. Previous \textit{Spitzer IRS} observations of this disk suggested that  water vapor is abundant in the innermost 2 au of the disk, while OH, CO$_2$, and HCN gas were only tentatively identified \citep{Salyk_2011}. Far-IR observations with \textit{Herschel}, on the other hand, have only placed an upper limit of 75.3 mK km/s for the integrated flux of the $1_{1,0} - 1_{0,1}$ water vapor line in the far-IR, and ro-vibrational lines were not detected either \citep{Banzatti_2023}.

\subsection{Data Acquisition and Reduction}

The observations presented in this paper were obtained using MIRI MRS as part of the JWST Cycle 1 program GO-2025 (P.I. Karin Öberg). AS~209 was observed starting on UTC August 2, 2022 01:19 adopting a point-source setting centered on (R.A. (J2000) = 16:49:15.2956, Decl. (J2000) = -14:22:09.01), with an integration time of 42 minutes in each of the three SHORT, MEDIUM, and LONG instrument sub-bands. The observations consist of six groups per integration, corresponding to the maximum possible before saturation due to the source brightness. Together, the four Integral Field Units (IFUs)---if observed in the three sub-bands---provide uninterrupted coverage from 4.9 to 28.1$\mu$m, with a resolving power of $R\sim3700$ in channel 1 ($4.9-7.6$ $\mu$m), $\sim2800$ in channel 2 ($7.5-11.7$ $\mu$m), $\sim2400$ in channel 3 ($11.5-18.0$ $\mu$m), and $\sim1500$ in channel 4 ($17.7-28.1$ $\mu$m).


The data was reduced using version 1.11.0 (CRDS context \texttt{jwst\_1105.pmap}) of the JWST pipeline \citep{Bushouse_2023}. For further details regarding the data reduction and 1D spectra extraction, we refer the reader to \citet{Pontoppidan_2023}. Being an atypically bright point-source as observed by MIRI MRS, the spectrum of AS~209 is severely affected by residual fringes, even after applying the fringe correction procedure \texttt{ResidualFringeStep} built into the JWST pipeline. We observe broad, quasi-periodic artifacts in all channels, and significant high-frequency variations in brightness versus wavelength around $10-12$ $\mu$m and in all of channel 4. These artifacts complicate the detection of real emission lines considerably, as well as the location of the true continuum baseline in each channel. We thus apply an empirical fringe correction using two asteroid calibrators from JWST GO program 1549 (P.I. Klaus Pontoppidan), which is also fully described by \citet{Pontoppidan_2023}.

\begin{figure*}
    \centering
    \includegraphics[width=0.98\textwidth]{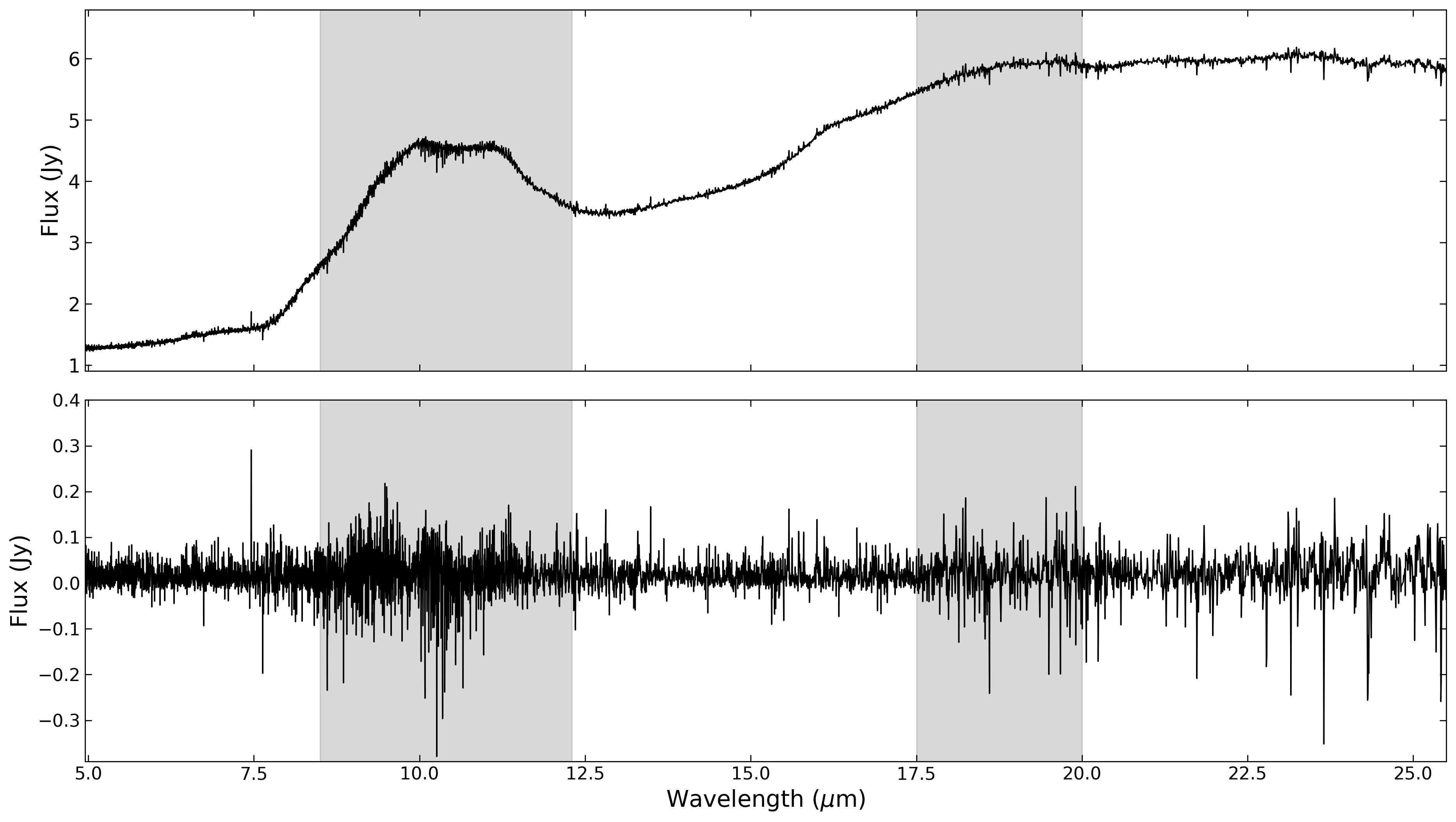}
    \caption{\textit{Top:} The complete MIRI MRS spectrum of the disk of AS~209. \textit{Bottom:} The continuum-subtracted spectrum, following the procedure described in Appendix \ref{ap:continuum}. The shaded regions in each panel show the wavelength ranges that were excluded from our analysis.}
    \label{fig:fullspec}
\end{figure*}

The top panel of Figure \ref{fig:fullspec} shows the full MIRI MRS spectrum of the AS~209 disk, and the bottom panel shows the continuum-subtracted spectrum. To subtract the continuum, we developed an iterative Gaussian Processes decomposition technique that allows us to fit a baseline to the data without having to identify all line-free regions amid the noisy background. The details are presented in Appendix \ref{ap:continuum}. Even after the correction for residual fringes, some regions of the spectrum, including most of channel 4, still exhibit multiple artifacts, bad pixels, and overall low SNR. We decide to exclude the 8.5 -- 12.3, 17.5 -- 20.0, and 25.5 -- 28.1 $\mu$m wavelength ranges (shaded regions in Figure \ref{fig:fullspec}) from our analysis. 

\begin{figure*}
    \centering
    \includegraphics[width=0.9\textwidth]{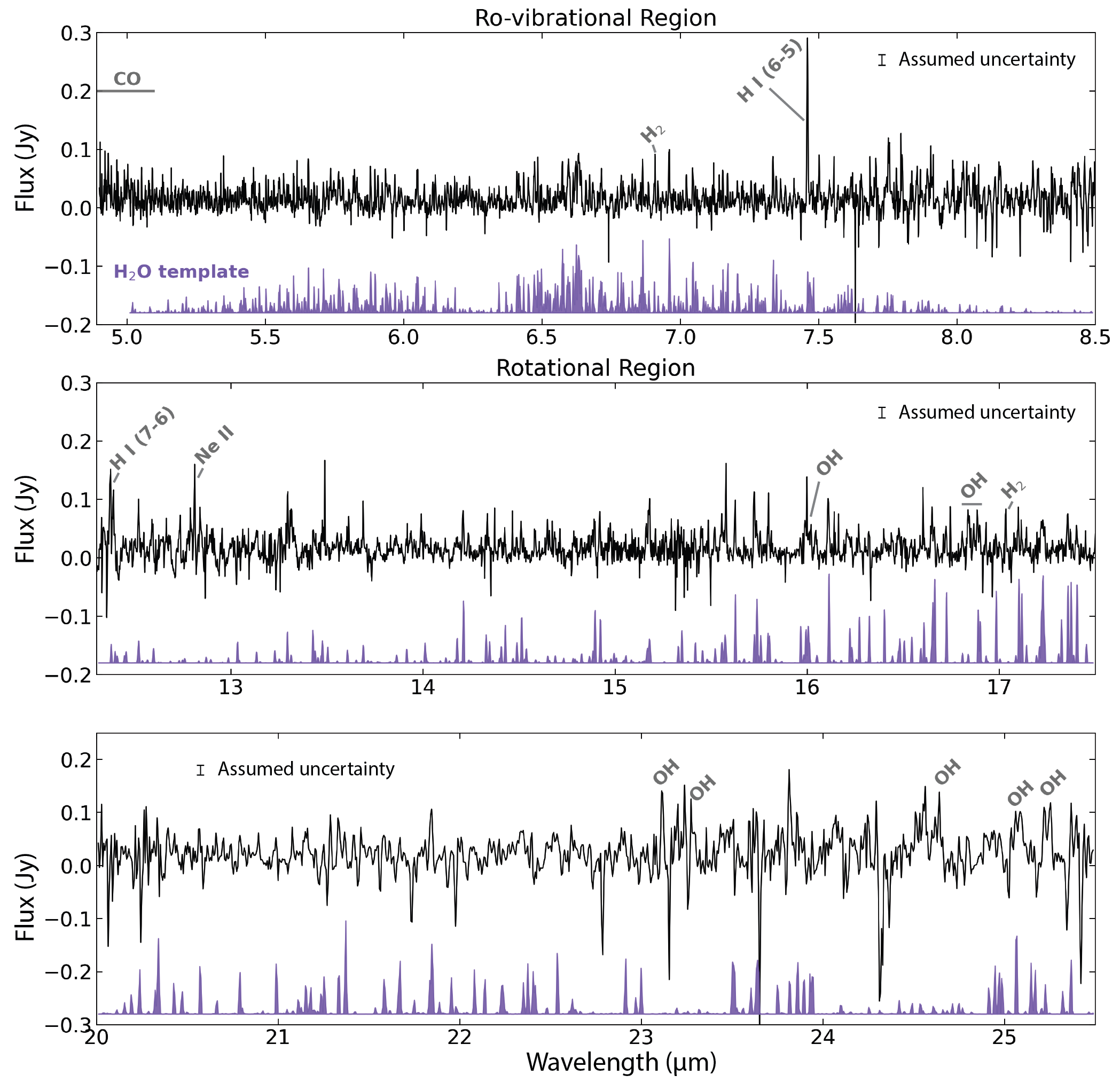}
    \caption{\textit{Top:} A closer view of the continuum-subtracted spectrum in the ro-vibrational (4.9 -- 8.5 $\mu$m) region.  A vertically-shifted template water vapor model manually scaled to match the data is shown in violet. \textit{Middle and bottom:} Same as top panel, for the rotational region of the spectrum. Note that the template model had to be scaled down by a larger amount in the ro-vibrational region compared to the rotational region. Emission lines from other species that could be visually identified are labeled in gray.}
    \label{fig:windows}
\end{figure*}

Figure \ref{fig:windows} provides a closer view at the regions of the MIRI MRS spectrum analyzed in this paper. In each panel, we include a template water vapor model (in violet) with an excitation temperature of 700 K and column density of 10$^{18}$ cm$^{-2}$, scaled `by-eye' to match the data. These values were chosen based on fits to high-energy water lines in other disks observed with MIRI MRS \citep{Banzatti_2023b}. As shown in the top panel, MIRI MRS reveals the presence of \textit{ro-vibrational} water emission in the disk of AS~209 for the first time, which is too weak to be detected from the ground \citep{Banzatti_2023}. We label some of the most prominent emission lines from other species that could be visually identified, namely CO, H$_2$, OH, and a few atomic lines. Based on visual inspection alone, we find no evidence of significant CO$_2$, HCN, nor hydrocarbon emission lines in the spectrum, in line with the observations provided by \textit{Spitzer IRS}. Since the 4.9 -- 8.5 $\mu$m region is dominated by the ro-vibrational bending mode of water, hereafter we refer to this wavelength range as the ro-vibrational region. The 12.3 -- 17.5 and 20.0 -- 25.5 $\mu$m wavelength ranges are dominated by rotational water emission, and we refer to these as the rotational region. 

\section{Analysis} \label{sec:analysis}

In the following we analyze the spectrum in detail to derive the column density $(N_{mol})$, excitation temperature $(T_{ex})$, and emitting area in the disk $(A_d)$ of the two species found based on visual inspection: H$_2$O and OH. We then perform a deeper search for emission of other molecules of interest: H$_2$, CO$_2$, C$_2$H$_2$, and HCN, and estimate their column densities when possible.

\subsection{Line Modeling}\label{sec:linemodel}

To simulate the molecular emission, we developed and deployed the GPU-accelerated LTE slab modeling Python package \texttt{iris} \citep{MirzaRomero_2023}. The line modeling prescription follows that presented by \citet{Tabone_2023}, adopting a detailed treatment of optical depth effects and line overlap and using spectroscopic information from the HITRAN database \citep{Gordon_2022}. In brief, we generate a high-resolution ($R \sim 10^5$) opacity-weighted flux model for each species, with an optical-depth grid described by
\begin{equation}
    \tau(\lambda) = \sum_{n=1}^N \tau_{0,n} \exp \left( \frac{- 4 \log(2)}{\Delta V^2} (\lambda - \lambda_{0,n})^2\right),
    \label{eq:taugrid}
\end{equation}
for each line $n$, where $\lambda_{0,n}$ is the line center wavelength and $\Delta V$ is the line full width at half maximum (FWHM). Here, $\tau_{0,n}$ is the line center optical depth
\begin{equation}
    \tau_{0,n} = \frac{A_{u,l} N_{mol} \lambda_{0,n}^3}{\Delta V} \Theta_{ul},
\end{equation}
with
\begin{equation}
    \Theta_{ul} = \left( \frac{\log 2}{16 \pi^3} \right)^{0.5} \left( x_{l} \frac{g_u}{g_l} - x_u\right),
\end{equation}
where $A_{u,l}$ is the transition Einstein coefficient, $x_{u}$ and $x_{l}$ are the upper- and lower-level populations, respectively, and $g_{u}$ and $g_{l}$ their degeneracies. The line FWHM is assumed to be the sum in quadrature of a fixed turbulent component $(v_{turb})$ and a thermal broadening component
\begin{equation}
    v_{therm} = \left(\frac{k_B T_{ex}}{\mu m_p}\right)^{0.5},
\end{equation}
where $\mu$ is the molecular weight of each species, $k_B$ is the Boltzmann constant, and $m_p$ the proton mass. As the observations are spectrally unresolved, the turbulent width cannot be directly determined, and we decide to use the same value assumed by \citet{Salyk_2011} and \citet{Tabone_2023}, $v_{turb} = 4.71 \text{ km/s}$ (2 km/s standard deviation of the Gaussian line profile). Since most of the lines are expected to be optically thick, we caution that the derived column densities scale roughly as $1/\Delta V$. Finally, the flux model is calculated as 
\begin{equation}
    F(\lambda) = \frac{A_d}{d^2} B_{\nu}(T_{ex}) (1 - e^{-\tau(\lambda)}),
\end{equation}
where $B_{\nu}(T_{ex})$ is the full Planck function and $d$ the distance to AS~209, assumed to be 121 pc.

The above prescription is computationally expensive---each species adds hundreds to thousands of transitions, all of which are modeled at high-resolution as described above---yet necessary given the plethora of emission lines that overlap within their intrinsic line widths. Ignoring these effects results in a severe over-estimation of peak line intensities, even at moderately high column densities ($\gtrsim 10^{18}$ cm$^{-2}$). We convolve the emission models with Gaussian kernels to approximate the resolving power of MIRI MRS ($R =$ 3400, 2300, 1700 for the 4.9 -- 8.5, 12.3 -- 17.5, and 20 -- 25.5 $\mu$m wavelength ranges, respectively) and downsample the resulting models to the MIRI MRS wavelength grid (while conserving the total flux) using the Python package \texttt{SpectRes} \citep{Carnall_2021} to compare with the data.  

\subsection{Bayesian Retrieval} \label{sec:bayes}

We model the ro-vibrational and rotational regions independently and include different species in each wavelength range of interest. For the rotational region, we include H$_2$O, OH, CO$_2$, C$_2$H$_2$, and HCN in the 12.3 -- 17.5 $\mu$m range, and only H$_2$O and OH in the 20 -- 25.5 $\mu$m range. Note that the contributions of H$_2$O and OH in the 12.3 -- 17.5 and 20 -- 25.5 $\mu$m ranges are fit together. For the ro-vibrational region, on the other hand, we only consider the contributions from H$_2$O based on the rotational region results (see Section \ref{sec:resultslow}). We emphasize that when multiple species (or multiple components for a single species) are included, all of them are modeled simultaneously, as modeling and subtracting each one individually would not correctly account for line opacity overlap along the line of sight. The rest of this section elaborates on the specific fitting details. 

We perform Bayesian inference in a hierarchical framework, using the Python Nested Sampling package \texttt{dynesty} \citep{Speagle_2020} to explore the posterior distribution of each parameter of interest. By iteratively adding molecules to the fit, we are able to better guide the sampler and avoid adding complexity to the model without sufficient evidence. To summarize the method described below, we add molecules to each emission model in succession, and each time a new molecule is included, we use the posterior distribution of the previous (simpler) fit to inform the prior distributions used in the new model. 

\subsubsection{Water Vapor}
For the rotational region, we begin by fitting an H$_2$O model using the following Uniform prior distributions:
\begin{equation}
\begin{split}
\log_{10}T_{ex} &= \mathcal{U} (2.17, 3.30) \text{ K} \\
\log_{10} N_{mol} &= \mathcal{U} (13, 22) \text{ cm}^{-2}\\
\log_{10} A_{d} &= \mathcal{U} (-3, 3) \text{ au}^{2},
\label{eq:uniform}
\end{split}
\end{equation}
with bounds informed by the previous observations and modeling efforts of water vapor emission in disks as discussed in Section \ref{sec:intro}. The same priors are used to model the H$_2$O emission in the rotational region. Note that the temperature bounds above correspond to 150 and 2000 K, and the log area bounds correspond to equivalent radii of 0.017 -- 17.8 au assuming circular emitting areas.  The nested sampling is performed using multiple ellipsoid decomposition and random walk sampling with $n_{live}$  = $10\times n_{dim}$ chains (i.e. live points), where $n_{dim}$ is the number of parameters in the model. We use a Normal likelihood assuming a constant 20 mJy uncertainty for all pixels\footnote{This is approximately 2$\times$ the standard deviation of pixels with negative flux values in the 20.0 -- 25.5 $\mu$m range, excluding bad pixels (see Appendix \ref{ap:continuum})}. The run is terminated once the estimated change in the remaining evidence $\Delta \log \mathcal{Z}$ satisfies 
$$
\Delta \log \mathcal{Z} \leq 0.001 (n_{live} - 1) + 0.01.
$$

\subsubsection{OH}

The next molecule added to the rotational region model is OH, which is also visually identified in the spectrum. To perform this fit, we begin by using the same Uniform prior distributions described by Equation \ref{eq:uniform} for OH, and Normal prior distributions for H$_2$O. In particular, we take the median and $3 \times$ the standard deviation of the posterior distribution of the previous H$_2$O fit to construct the new Normal priors. We choose to use $3 \times$ the standard deviation of the previous results to avoid under-estimating the parameter uncertainties in the new fits. The number of live points is increased and the run is terminated using the same criteria as described above. 

We find that the above fit produces multi-modal posterior distributions for OH, with three main solutions: one with $T_{ex}$ of $\sim 400$ K, one near 700 K, and a third one at approximately 1500 K. The first two solutions have similar emitting areas of $\sim 1 $ au$^2$ and column densities of order 10$^{18}$ cm$^{-2}$, while the third one requires a column density $< 10^{16}$  cm$^{-2}$. This is not unexpected, since the gas responsible for inner disk OH emission is thought to be in non-LTE, and slab models with multiple excitation temperature components are typically required to reproduce the observed line ratios \citep[e.g.][]{Najita_2010, Banzatti_2012, Carr_2014}. Of the three solutions, we focus on the one with $T_{ex}$ $\approx 700$ K (see Section \ref{sec:results}) for two main reasons: First, this solution has the highest likelihood and can be considered to be the global minimum in the parameter space allowed by the priors. Second, we find that the low temperature solution misses the high-energy OH lines seen at 16.8 $\mu$m, while the high temperature solution cannot reproduce the lower energy lines beyond 20 $\mu$m. The $700$ K solution on the other hand provides the best overall fit to the data. We also experimented with fitting multiple OH components simultaneously, but the models did not converge.

Based on the findings described above we repeat the H$_2$O + OH fit, this time using a Normal $T_{ex}$ prior for OH centered at 700 K with standard deviation of 200 K, and the same $N_{mol}$ and $A_d$ Uniform priors. With the more restrictive temperature prior we no longer recover the multi-modal solutions for OH, and thus the posterior distribution provides a better estimate of the uncertainty for the main solution.

\subsubsection{A Cold Water Vapor Component}

The fit described before results in a hot and compact water emission component, as further detailed in Section \ref{sec:resultslow}. After careful inspection of the residuals, we find minor excess emission features that may be associated with low-energy H$_2$O lines around 21.2, 21.8, 23.8, and 25.5 $\mu$m, which are not captured by the best fit water model. As mentioned in Section \ref{sec:intro}, mid-IR rotational water vapor spectra in disks tends to exhibit excitation temperature gradients, sometimes associated with multiple reservoirs produced by distinct water vapor formation mechanisms. We thus explore whether there could be a colder water reservoir contributing to the spectrum.

We fix the temperature of this second component to 400 K, as is observed in other T-Tauri systems \citep{Banzatti_2023b}, and fit only for the emitting area and column density using the same uniform priors as described above. We also experiment by allowing the excitation temperature to vary, and get a solution with $T_{ex} \sim 200 $ K. However, given the low quality of the data in channel 4, we decide not to derive any conclusions regarding the temperature of cold water vapor from the long wavelength emission, and only use the fit to obtain an upper limit for the amount of cold water vapor. Similarly, we caution against interpreting the best fit cold H$_2$O emitting area as a snowline location tracer, given that it is strongly correlated to the choice of $T_{ex}$.

\subsubsection{Other Molecules}

For the rest of the species included in the 12.3 -- 17.5 $\mu$m range, CO$_2$, C$_2$H$_2$, and HCN, we are unable to fit all $N_{mol}$, $T_{ex}$, $A_d$, likely due to a mix of intrinsically low fluxes, few observable lines, and poor SNR. Specifically, their column densities and emitting areas are degenerate, and we cannot place any meaningful constraints on their excitation temperatures. Thus, we attempt to estimate only the column density of each species, using the following uniform prior:
$$
\log_{10} N_{mol} = \mathcal{U} (12, 20) \text{ cm}^{-2}.
$$ 
We use the same area and excitation temperature obtained for the hot water vapor component to estimate $N_{mol}$ for CO$_2$, HCN, and C$_2$H$_2$, which are added to the model in that order. Note that when adding each of these species we no longer let the parameters of the previous models vary, but rather fix them using the median of their final posterior distribution. At the expected abundances, these species emit most prominently in non-overlapping wavelength regions (see Figure \ref{fig:rot1}), and thus should not affect the retrieved properties of each other, nor those of H$_2$O and OH.

\section{Results} \label{sec:results}

\subsection{Rotational Region} \label{sec:resultslow}

\begin{figure}
    \centering
    \includegraphics[width=0.45\textwidth]{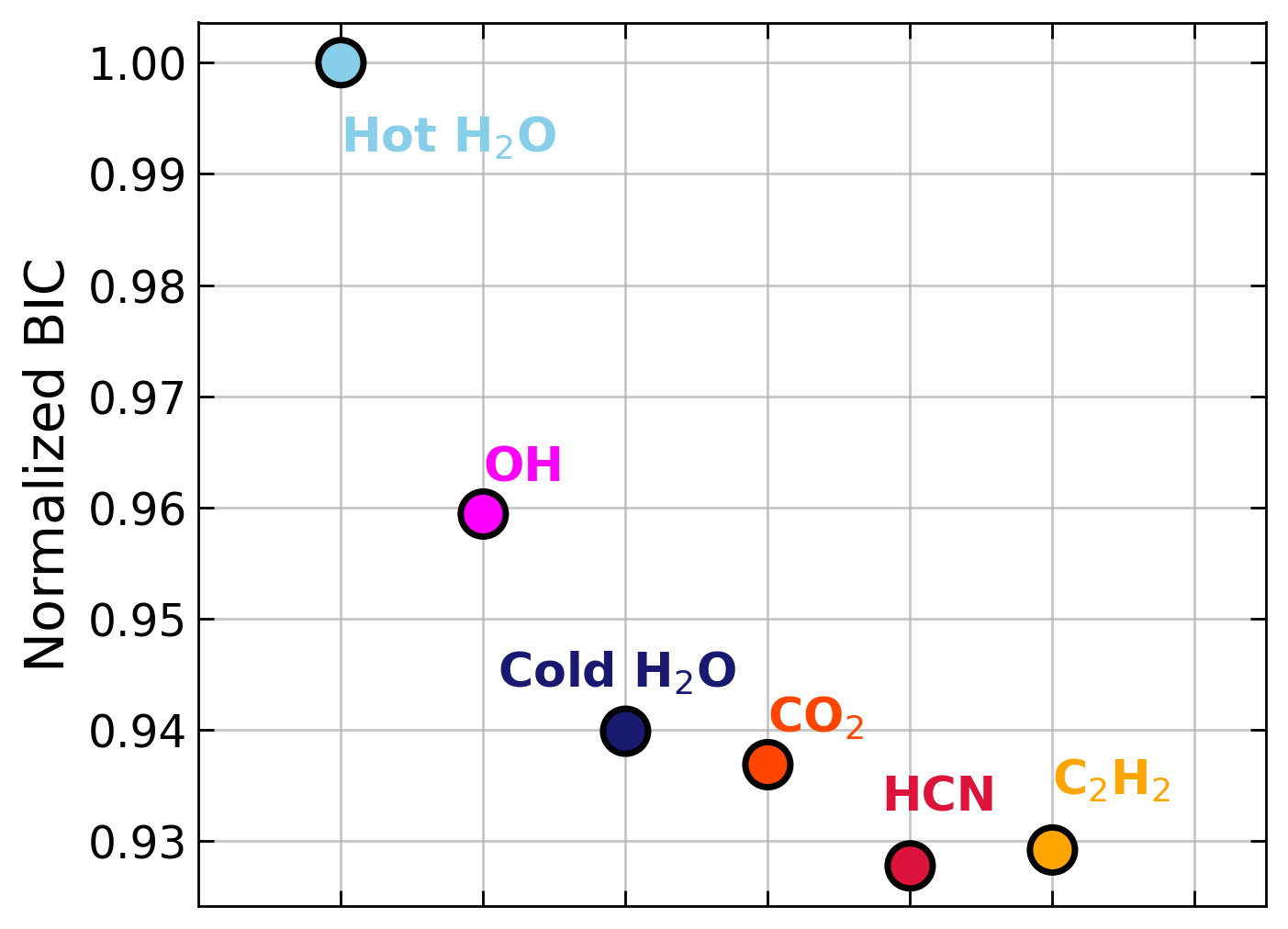}
    \caption{Bayesian Information Criterion (BIC) obtained by adding each species to the rotational region flux model. The values have been normalized for clarity.}
    \label{fig:bic}
\end{figure}

As described in Section \ref{sec:analysis}, we attempt to fit the emission of hot and cold H$_2$O, OH, CO$_2$, HCN, and C$_2$H$_2$. Since we can only visually confirm the presence of hot water vapor and OH in the spectrum, we emphasize that CO$_2$, C$_2$H$_2$, HCN, and cold H$_2$O can be considered, at best, marginally detected. As a way to decide whether a species is detected, marginally detected, or non-detected, we use the Bayesian Information Criterion (BIC), which penalizes the increased complexity of each model even if it has a higher likelihood:
$$
\text{BIC} = n_{dim} \log k - 2 \log \mathcal{L}_{\star},
$$
where $n_{dim}$ is the dimensionality of the problem, $k$ is the total number of observations, and $\mathcal{L}_{\star}$ the log-likelihood of the model evaluated using the median of the posterior distribution for each parameter. Specifically, hot water vapor is considered an unambiguous detection, and the rest of the molecules are:

\begin{enumerate}
    \item \textit{Detected}: if we can identify multiple lines, the strongest component lies at the $> 3\sigma$ level after subtracting hot H$_2$O from the data, and the BIC decreases by including the species in the model.
    \item \textit{Marginally detected}: if it cannot be unambiguously identified in the data, but the BIC decreases by including it in the model.
    \item \textit{Non-detected}: if it cannot be unambiguously identified in the data, and the BIC does not decrease by including it in the model. 
\end{enumerate}
Figure \ref{fig:bic} shows how the BIC is modified by adding additional species to the model, starting from hot water vapor. OH is detected, CO$_2$, HCN, and cold H$_2$O are marginally detected, and C$_2$H$_2$ is not detected. CO and H$_2$ are considered detections too, but are not included in the excitation analysis.

\begin{figure*}
    \centering
    \includegraphics[width=0.98\textwidth]{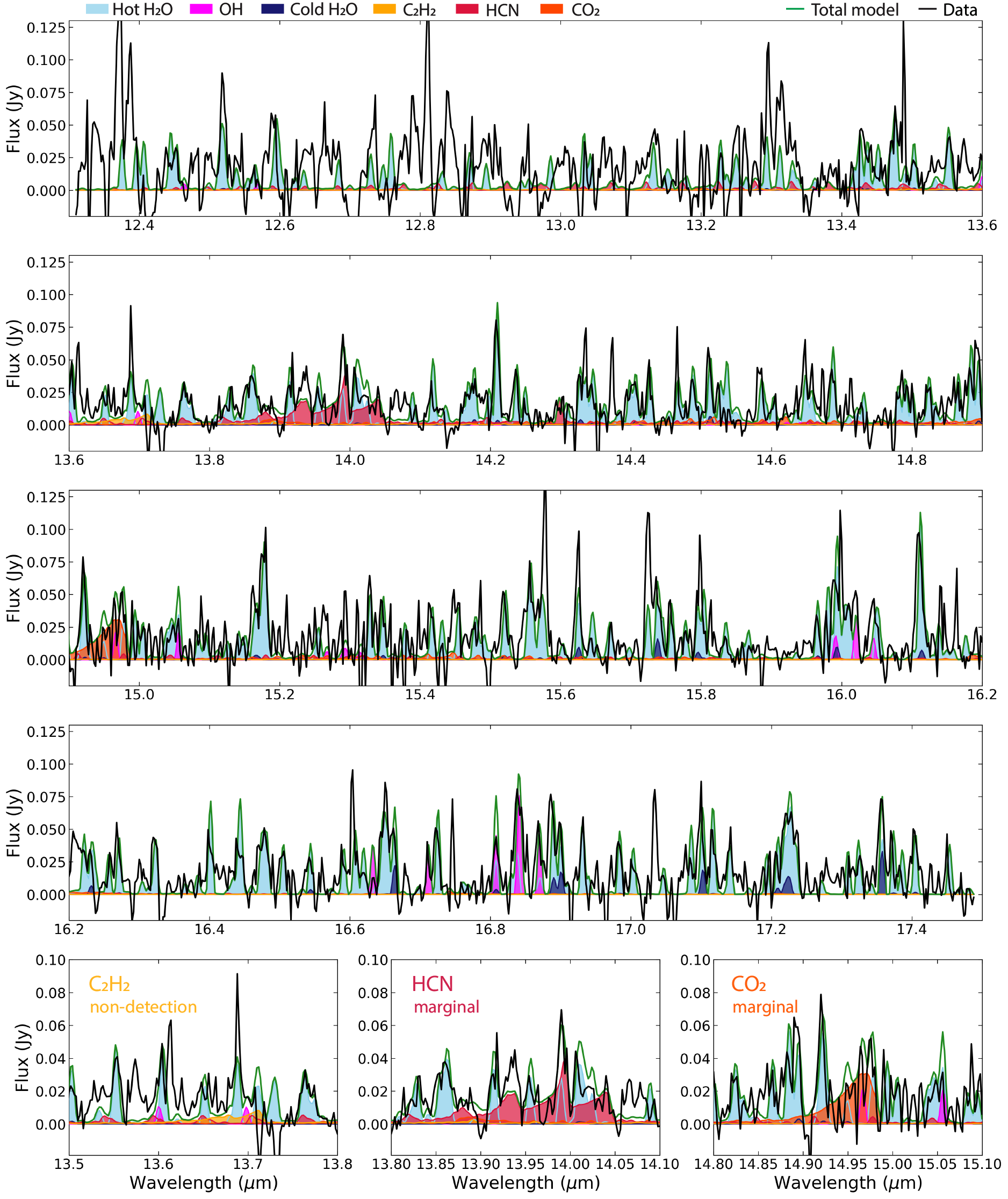}
    \caption{\textit{Top four panels:} The best fit model (green line) compared to the MIRI MRS spectrum (black line) in the 12.3 -- 17.5 $\mu$m range. Individual contributions from each species are shown as shaded areas. \textit{Bottom:} A closer view of the brightest C$_2$H$_2$, HCN, and CO$_2$ branches. Note that the best-fit C$_2$H$_2$, HCN, and CO$_2$ models shown are interpreted as upper limits.}
    \label{fig:rot1}
\end{figure*}

\begin{figure*}
    \centering
    \includegraphics[width=0.98\textwidth]{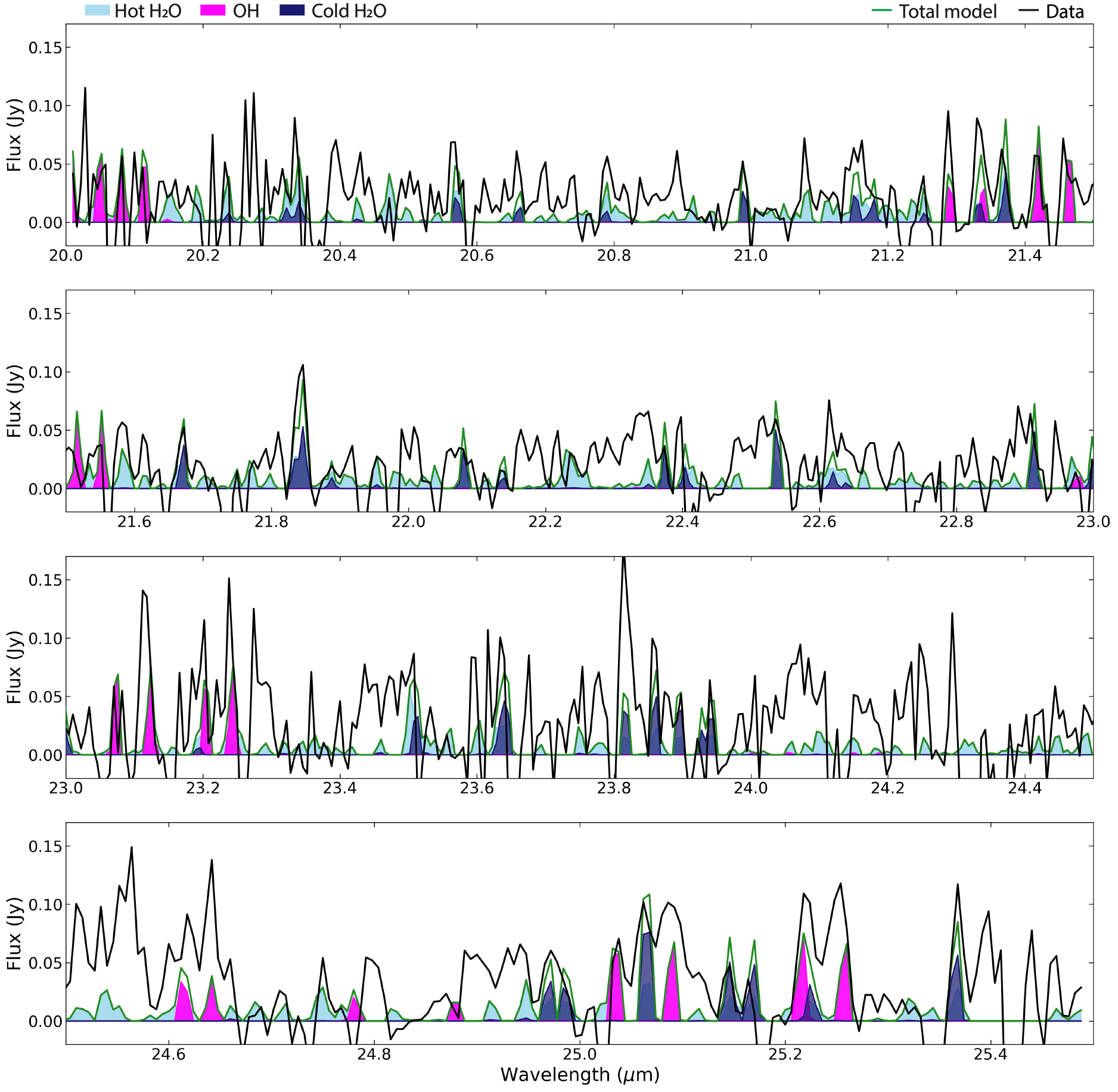}
    \caption{Similar as Figure \ref{fig:rot1} for the 20.0 -- 25.5 $\mu$m range.}
    \label{fig:rot2}
\end{figure*}

Table \ref{tab:bestfit} lists the best-fitting values and confidence intervals for each parameter of interest in the final fit. The best-fit values are chosen to be the median of the posterior distributions, while the uncertainties correspond to the 0.0015 and 0.9985 percentiles ($3 \sigma$ confidence intervals). We caution that the reported uncertainties reflect the assumed flux uncertainty of 20 mJy, but likely under-estimate the true parameter uncertainties. This is because the likelihood function and retrieval framework used in this work can only include Normally-distributed noise, but does not account for correlated noise nor the uncertainty related to both the absolute flux calibration and continuum baseline. Based on visual inspection of the posteriors, we do not notice any strong correlations between parameters. From here onward, the emission models referred to as `best-fits' correspond to those evaluated using the median of the posterior distribution for each parameter. Since CO$_2$ and organics are only marginally/not detected, we consider their best-fit column density to be upper limits. In the case of cold H$_2$O for which we can fit both emitting area and column density, we consider the observable mass (product of best-fit column density and emitting area) to be an upper limit as well. Table \ref{tab:bestfit} further lists the inferred total observable mass of each molecule, and their ratio relative to hot H$_2$O.  

\begin{deluxetable*}{llllllll}
\label{tab:bestfit}
\tabletypesize{\footnotesize}
\tablecolumns{8}
\tablewidth{0pt}
\tablecaption{ Best-Fit Model Details \label{table:results1}}
\tablehead{
\colhead{Region} & \colhead{Species} & \colhead{Status} & \colhead{$T_{ex}$ (K)} & \colhead{$N_{mol}$ (cm$^{-2}$)} & \colhead{$A_d$ (au$^2$)} & \colhead{$M$ ($M_\earth$)} & \colhead{$M[X] / M[$Hot H$_2$O$]$}}
\startdata 
Rotational & & & & & & & \\
& Hot H$_2$O & Detected & $830^{+110}_{-90}$ & $2.8^{+4.7}_{-1.7} \times 10^{19}$ & $5.8^{+1.5}_{-1.1} \times 10^{-2}$ & $1.8 \times 10^{-6}$ & -- \\
& OH & Detected & $730^{+70}_{-70}$ & $1.7^{+1.6}_{-0.9} \times 10^{18}$ & $2.9^{+0.9}_{-0.8}\times 10^{-1}$ & $5.4 \times 10^{-7}$ & 0.3 \\
& Cold H$_2$O & Marginal & [400] & $4.6 \times 10^{17}$ & $8.0 \times 10^{-1}$  & $< 4.1 \times 10^{-7}$ & $<0.2$ \\ 
& CO$_2$ & Marginal & [830] & $< 5.5 \times 10^{16}$ & [$5.8 \times 10^{-2}$] & $< 8.8 \times 10^{-9}$ & $<4.9 \times 10^{-3}$ \\
& HCN & Marginal & [830]  & $< 6.8 \times 10^{16}$ & [$5.8 \times 10^{-2}$] & $< 6.6 \times 10^{-9}$ & $<3.7 \times 10^{-3}$ \\
& C$_2$H$_2$ & Not Detected & [830] & $< 5.3 \times 10^{15}$ & [$5.8 \times 10^{-2}$] & $< 5.0 \times 10^{-10}$ & $<2.8 \times 10^{-4}$  \\
Ro-vibrational & & & & \\
& Ro-vib H$_2$O & Detected & $1200^{+70}_{-90}$ & $3.9^{+1.1}_{-0.8} \times10^{18}$ & $9.1^{+1.5}_{-0.9} \times10^{-3}$ & -- & -- \vspace{0.1cm}\\
\enddata
\tablecomments{\textit{From left to right}: Spectral region, species name, detection status, excitation temperature, column density, emitting area, observable mass, and mass ratio with respect to hot water vapor. Best-fit parameters correspond to the median of the posterior distributions. Confidence intervals correspond to the 0.0015 and 0.9985 percentiles. Brackets indicate parameters that were kept fixed during the fit. Observable masses are the product of the best-fit column density and emitting area, times the molecular weight of each species. We do not present a mass estimate for ro-vibrational H$_2$O, since the emission is affected by non-LTE excitation effects.}
\end{deluxetable*}

Figure \ref{fig:rot1} shows the best-fit model in the 12.3 -- 17.5 $\mu$m region. For clarity, the spectrum and model have been broken up into four vertically-stacked panels. Besides the best fit model, we show the individual contribution from each species as shaded areas of different colors, and provide additional panels with a zoomed-in view of the main organic bands. We find that the best fit hot H$_2$O model does an excellent job at reproducing most of the emission observed in this spectral region, except in the noisier 12.3 -- 13.6 $\mu$m region. Similarly, the 700 K OH model can reproduce the strongest high-energy OH lines near 16.8 $\mu$m and even those blended with water near 16.0 $\mu$m. CO$_2$ and the organics, if present, are completely blended with water. Yet we note that adding HCN and CO$_2$, which are marginally detected, improve the quality of the overall fit.

Finally, Figure \ref{fig:rot2} shows the best-fit model in the 20.0 -- 25.5 $\mu$m region. In this region, notice that hot H$_2$O has only minor contributions, and the data is dominated by noise. We can yet observe that the OH model matches the brightest lines near 21.1 and 21.5 $\mu$m well, but under-predicts the observed fluxes of lower-energy lines beyond 23 $\mu$m by about 50$\%$. Even with the higher noise, we notice that several bright features can be partially reproduced by the marginally-detected cold H$_2$O component, particularly at 21.4, 21.85, between 23.8 -- 24.0, and between 25.0 -- 25.4 $\mu$m.

\subsection{Ro-vibrational Region} \label{sec:resultshigh}

Here we present the results from fitting the water vapor emission in the ro-vibrational region (4.9 -- 8.5 $\mu$m range). However, we emphasize that emission from the ro-vibrational bending mode of water is generally found to not be in local thermodynamic equilibrium (LTE)---hence why it is fit separately from the rest of the spectrum---and the derived properties using LTE slabs may not be good estimates of the true gas temperature, column density, and extent. In fact, it is likely that the ro-vibrational water emission traces a similar, perhaps slightly hotter water vapor reservoir seen in the rotational region---but is simply suppressed due to excitation effects \citep{Banzatti_2023, Bosman_2022}.  

\begin{figure*}
    \centering
    \includegraphics[width=0.98\textwidth]{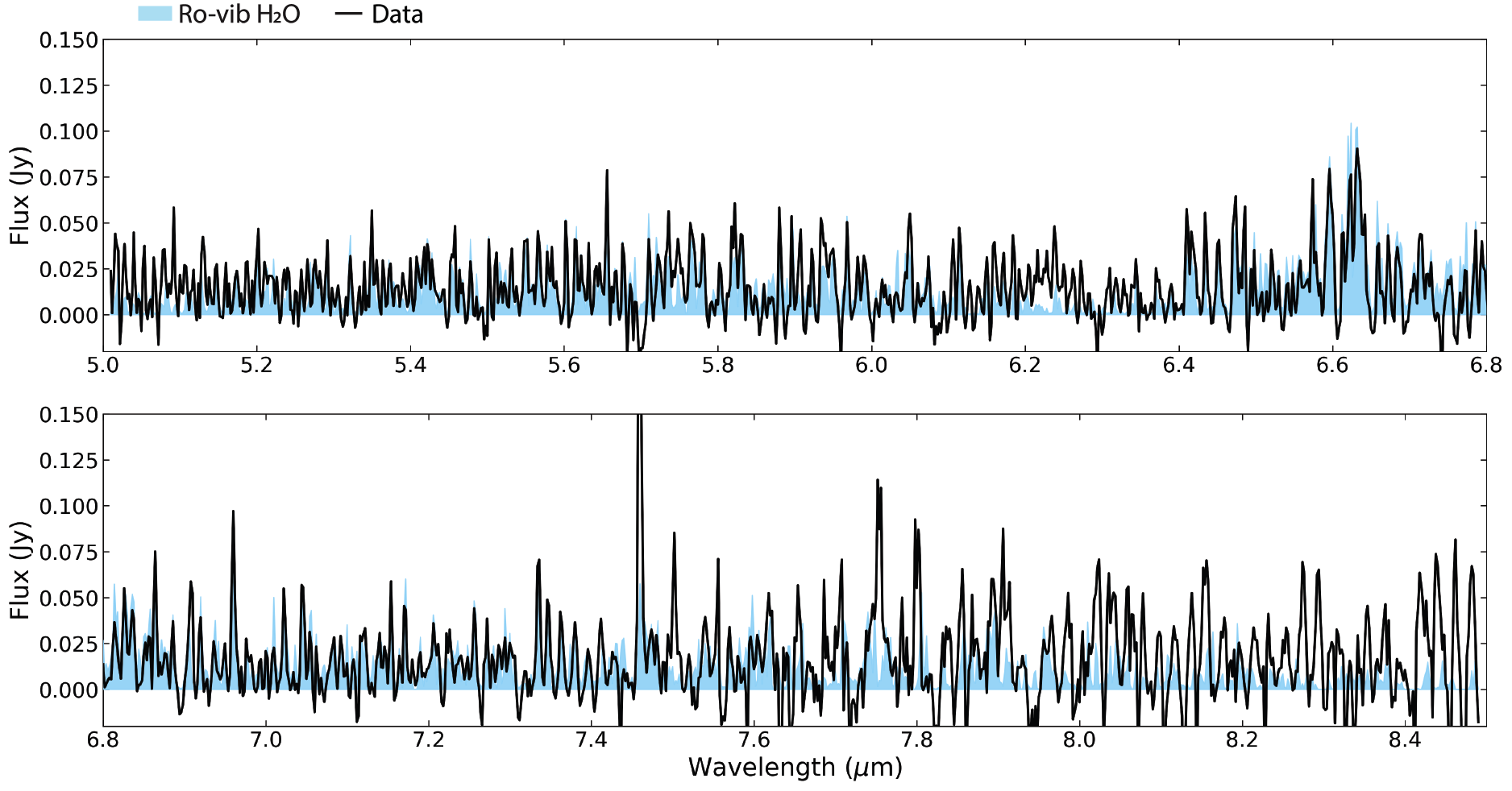}
    \caption{The best fit water vapor model (blue) compared to the data (black) in the ro-vibrational region.}
    \label{fig:rovib}
\end{figure*}

Figure \ref{fig:rovib} shows the best-fit model in the ro-vibrational region, both compared to the data and individually. We find that the ro-vibrational water vapor emission component is best modelled by a slab with area of $9 \times 10^{-3}$ au$^{2}$, a higher excitation temperature of 1230 K, and a lower column density compared to that traced by the rotational component, of $4 \times 10^{18}$ cm$^{-2}$. Note that while this H$_2$O model appears to be a good fit to the data, it severely under-predicts the observed water line fluxes in the rotational region. Similarly, the hot and cold water components fit to the rotational regions over-predict the observed water spectrum between 4.9 and 8.5 $\mu$m. The two regions cannot be reproduced by a single slab or sum of slab models. Based on the best-fit column densities of the other species in the rotational region, we do not expect any of them to be detected between 4.9 and 8.5 $\mu$m. 

\section{Discussion} \label{sec:discussion}

\subsection{The Mid-IR Spectrum of AS 209 in Context} \label{sec:incontext}


To guide the interpretation of our results, we carry out a detailed comparison between the MIRI MRS spectrum of AS 209 and those of two other systems: CI Tau and GK Tau. These provide some of the highest quality MIRI MRS data published for T-Tauri systems to date, and let us explore 1) if the properties of the hot water vapor reservoir in AS 209 are atypical, and 2) whether the weaker molecular features tentatively found in the AS 209 spectrum are consistent, within the noise, with those detected in other disk spectra. Furthermore, we choose these two systems to contrast the H$_2$O and OH excitation conditions in AS 209 to those of a similarly extended and structured disk (CI Tau, with sub-mm dust radius $R_{dust} = 190$ au) and a significantly more compact disk (GK Tau, $R_{dust} = 13$ au) \citep{Long_2019}. 

The GK Tau and CI Tau MIRI MRS spectra were obtained in program JWST Cycle 1 program GO-1640 (P.I. Andrea Banzatti), and were recently presented by \citet{Banzatti_2023b} in the context of water delivery. The spectra were reduced and de-fringed using the same calibration pipeline applied in this work to AS 209 \citep{Pontoppidan_2023}. To accurately compare all three spectra, we first scale the GK Tau and CI Tau data to the same distance as AS 209 (121 pc), assuming a distance of 160 pc for CI Tau and 129 pc for GK Tau \citep{Gaia}. Next, we use the bright and optically-thick high-energy H$_2$O lines near 16 $\mu$m to remove the effects of differing line emitting areas \citep{Banzatti_2020}. Specifically, we take the 16.05 -- 16.6 and 17.0 -- 17.3 $\mu$m spectral ranges, and find the scaling factor that minimizes the sum of squared differences between each spectrum and that of AS 209. CI Tau is scaled by a factor of 0.44, while the GK Tau spectrum does not require any additional scaling. 

\begin{figure}
    \centering
    \includegraphics[width=0.4\textwidth]{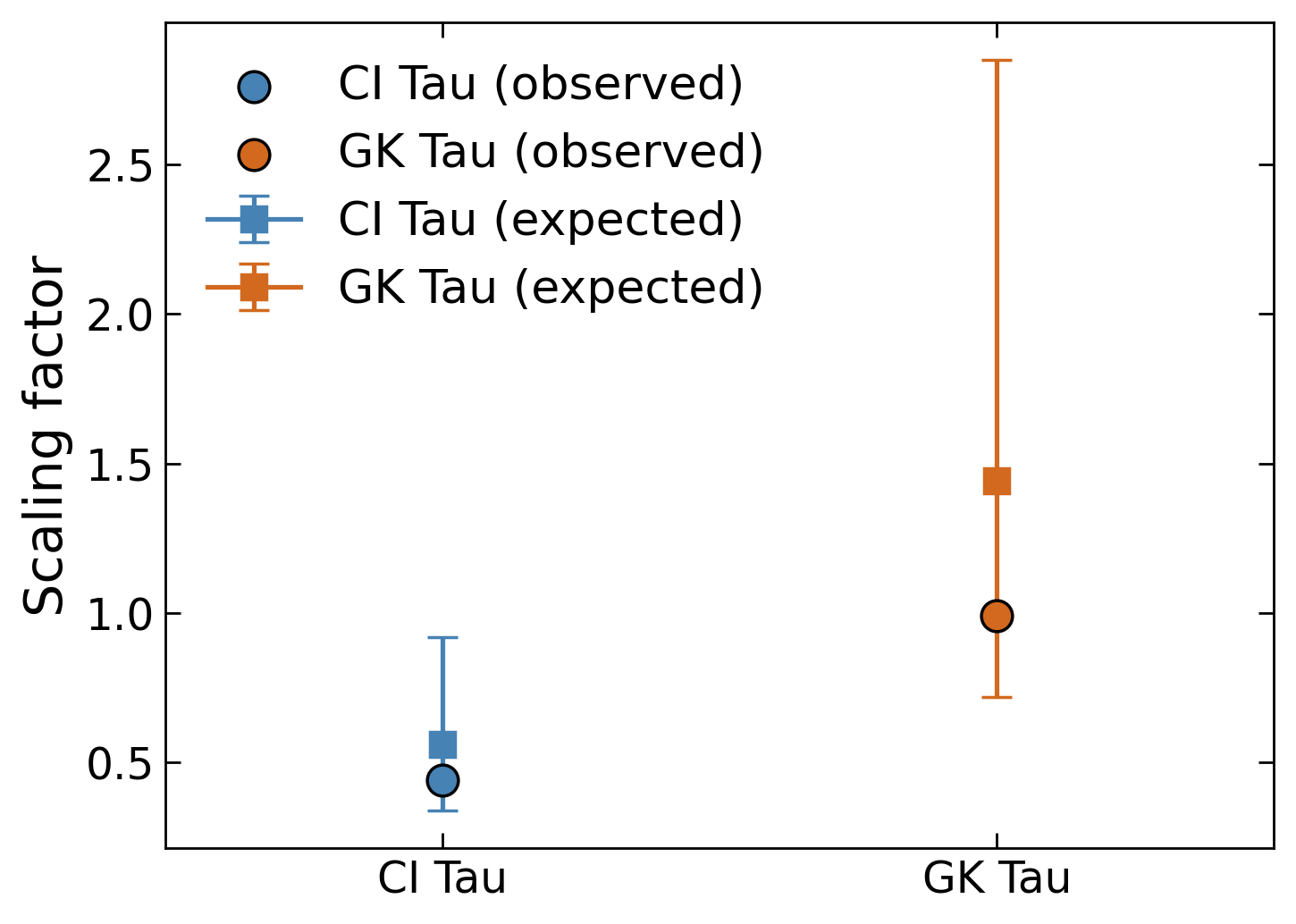}
    \caption{The scaling factors (circles) necessary to match the optically-thick high-energy H$_2$O line fluxes in GK Tau and CI Tau with those of AS 209. The squares and error bars show expected values and their uncertainty based on the $L_{acc}$ of each host.}
    \label{fig:scalings}
\end{figure}

A first test is to determine whether the scaling factors necessary to match the optically-thick H$_2$O line luminosities are consistent with the expectation based on the stellar accretion luminosity $(L_{acc})$ of each system. In particular, \citet{Banzatti_2023b} find that the optically-thick H$_2$O log line fluxes follow a relatively tight linear correlation with $\log_{10} L_{acc}$, with slope $\sim 0.59$. Figure \ref{fig:scalings} shows the observed and expected scaling factors, assuming $\log_{10} L_{acc}$ values of -1.12, -0.7, and -1.38 for AS 209, CI Tau, and GK Tau, respectively, and a typical $20\%$ uncertainty in $\log_{10} L_{acc}$ \citep{Fang_2018}. While the observed scaling factors are slightly lower than expected from $L_{acc}$ effects alone, they are still fully consistent within the uncertainty. That is, we find no evidence that the hot water vapor reservoir of AS 209 is significantly more compact or diffuse relative to these CTTs.

\begin{figure*}
    \centering
    \includegraphics[width=0.95\textwidth]{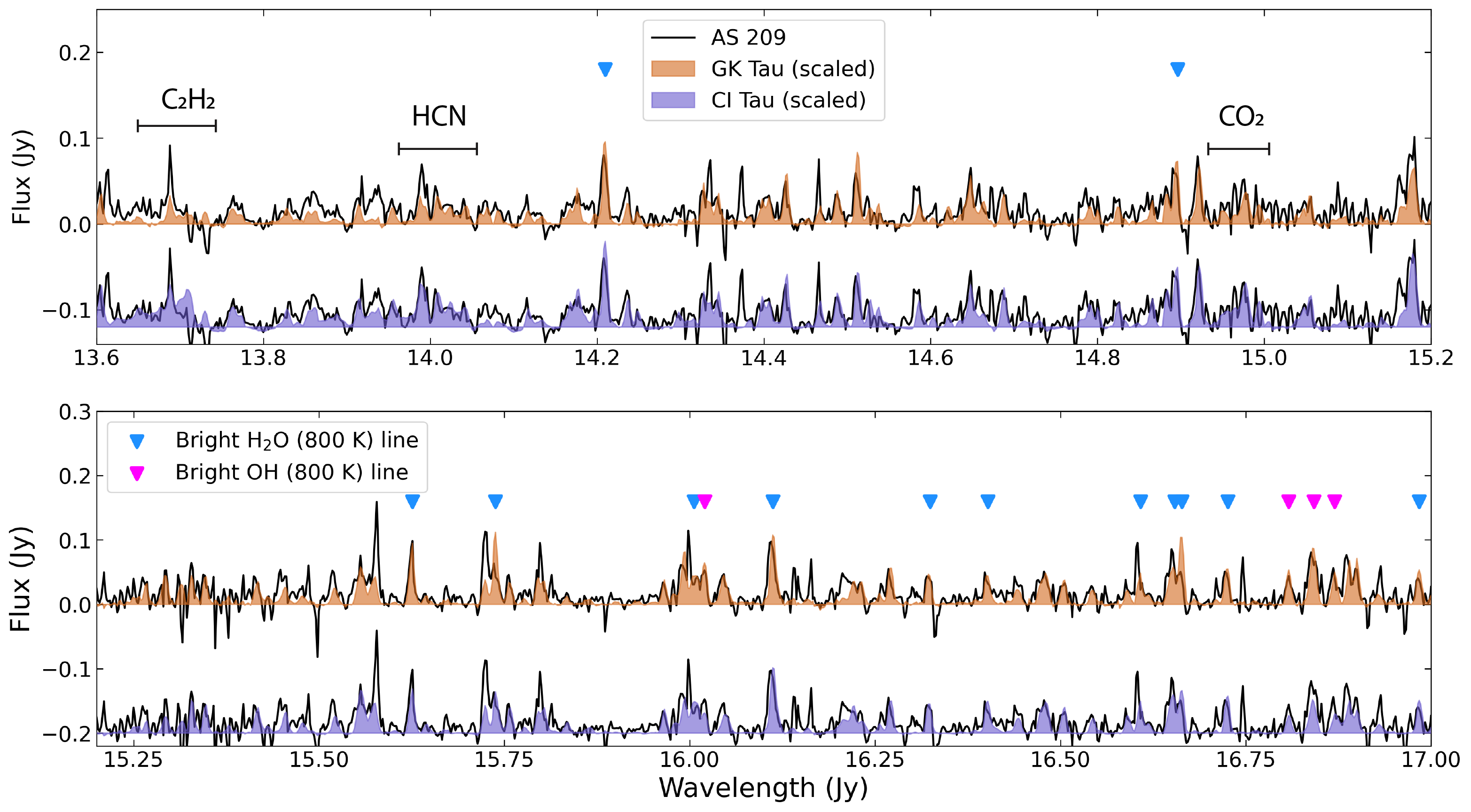}
    \caption{Comparison between the AS 209 spectrum and those of CI Tau and GK Tau, after scaling the latter to match the optically thick water vapor lines near 16 $\mu$m. The AS 209 spectrum is shown twice for each comparison and is shifted vertically for clarity.}
    \label{fig:comparedisks}
\end{figure*}

Note however that the retrieved emitting area and column density of hot H$_2$O in AS 209 are roughly an order of magnitude smaller and higher, respectively, compared to those reported for GK Tau and CI Tau by \citet{Banzatti_2023b}. We have reanalyzed the spectra of the latter two using the updated line modeling framework presented in this paper, and conclude that the difference does not reflect a real discrepancy in the water vapor distribution, but rather arises from a) a different intrinsic line FWHM used (1 km/s v. 4.7 km/s used in this work), and b) the distinct treatment of line overlap and opacities. These new water vapor retrievals will be presented in a forthcoming paper (Muñoz-Romero et al. in prep.).

Figure \ref{fig:comparedisks} shows the comparison between the MIRI MRS spectrum of AS 209 and the scaled spectra of CI Tau and GK Tau, from 13.6 to 17 $\mu$m. For both disks, we find that their spectra are remarkably similar to that of AS 209. The OH, HCN and CO$_2$ features detected in both CI Tau and GK Tau are consistent with the AS 209 data. In fact, based on visual inspection alone, the main HCN and CO$_2$ branches seem to be slightly brighter in AS 209 compared to GK Tau. We surmise that neither species is atypically depleted in AS 209, but rather they are only marginally detected due to the lower data SNR. On the other hand, it appears that C$_2$H$_2$ may indeed be suppressed in AS 209, as indicated by our best fit model as well, but deeper observations will be necessary to confirm this and any possible implications regarding the gas phase C/O ratio in the inner disk.

\subsection{Emission Variability}

We conclude by comparing the MIRI MRS spectrum of AS~209 and best fit models to the previous observations of this disk with \textit{Spitzer IRS}. AS~209 was observed as part of the cores to disks (c2d) \textit{Spitzer} Legacy program \citep{Evans_2003}, providing coverage from 9 to 37 $\mu$m. The inner disk molecular emission was subsequently presented and characterized by \citet{Pontoppidan_2010} and \citet{Salyk_2011}. In particular, \citet{Salyk_2011} reported the detection of water vapor and CO gas in AS~209, as well as a tentative detection of CO$_2$. For H$_2$O, \citet{Salyk_2011} reported a best-fit column density of $10^{20.9}$ cm$^{-2}$, an excitation temperature of 250 K, and an emitting area of 9 au$^2$, which was obtained by fitting the integrated fluxes of 65 individual water lines. We find that this model strongly over-predicts the water line emission seen in the MIRI MRS spectrum. Previous observations and modeling thus missed the hot water reservoir that dominates the new observations and inferred a much larger and colder water reservoir than we find using JWST.

\begin{figure*}
    \centering
    \includegraphics[width=0.95\textwidth]{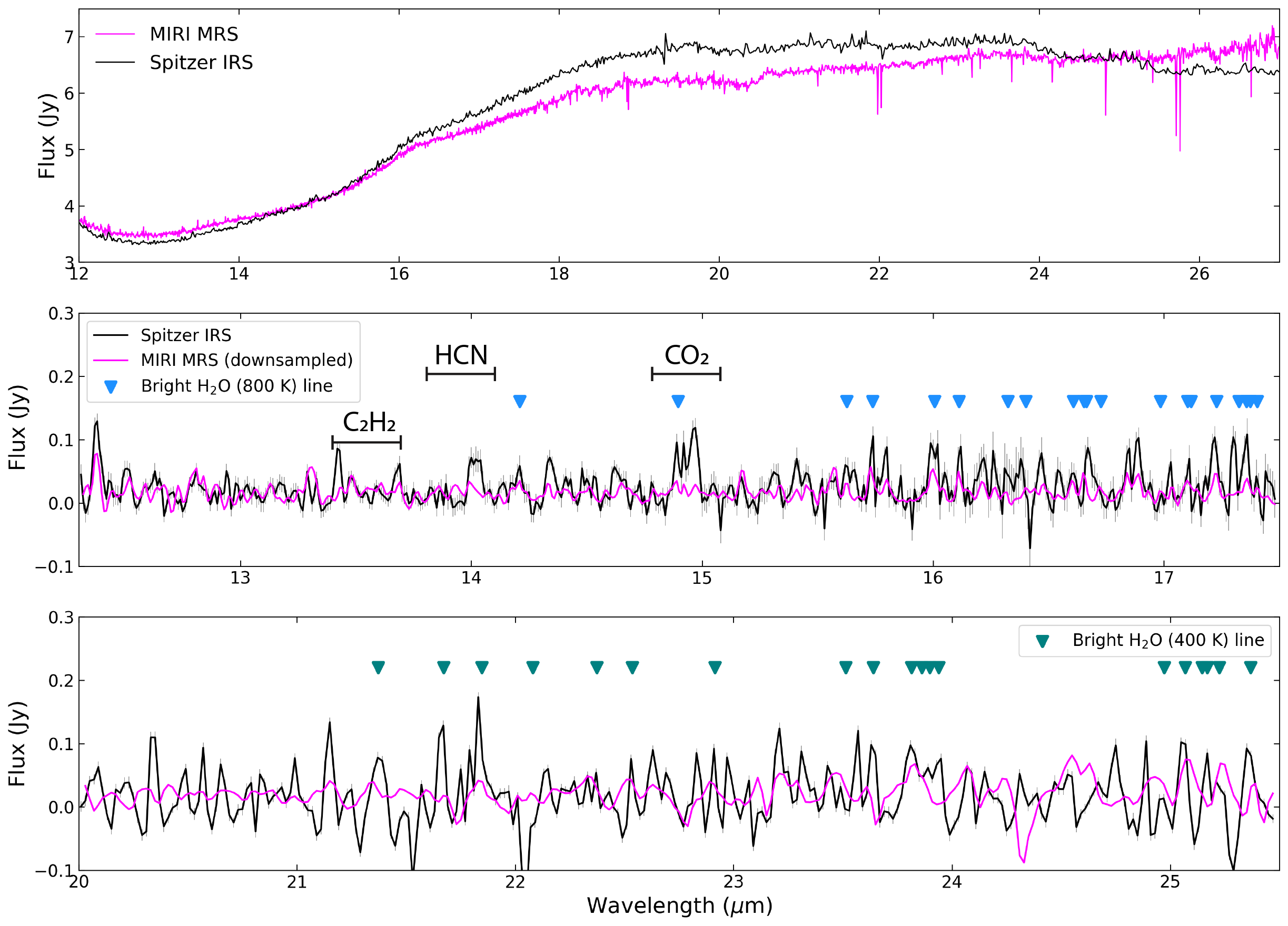}
    \caption{\textit{Top}: The MIRI MRS and \textit{Spitzer IRS} spectra of AS 209. \textit{Bottom two panels:} Continuum-subtracted \textit{Spitzer IRS} and MIRI MRS spectra, convolved down to $R \approx 600$. The expected location of the brightest hot (800 K) and cold (400 K) water vapor lines, as well as the brightest CO$_2$ and organic branches are indicated.}
    \label{fig:miri_v_irs_compact}
\end{figure*}

Figure \ref{fig:miri_v_irs_compact} shows the \textit{Spitzer IRS} and MIRI MRS (downsampled and convolved to the \textit{IRS} resolving power of 600) spectra of AS 209. We subtract the continuum baseline from the \textit{IRS} spectrum using the same pipeline applied to the MIRI MRS data. The error bars in the \textit{Spitzer IRS} spectrum correspond to the flux uncertainty produced by the reduction pipeline used by \citet{Pontoppidan_2010}. Note that the continuum emission at both epochs is quite similar, with the \textit{Spitzer IRS} continuum being at most $10\%$ brighter near 20 $\mu$m. On the other hand, we find substantial differences between the continuum-subtracted spectra. While some spectral regions, such as between 12 -- 14 $\mu$m are well matched, others require the MIRI MRS spectrum to be scaled up by factors of 2 -- 4 to match the \textit{Spitzer IRS} data. For instance, the region near the brightest CO$_2$ branch at 15 $\mu$m appears to be 4$\times$ as bright in  \textit{Spitzer IRS}, while some features between 16 -- 17 $\mu$m are twice as bright in \textit{Spitzer IRS.}

\begin{deluxetable}{lll}
\label{tab:factors}
\tabletypesize{\footnotesize}
\tablecolumns{3}
\tablewidth{0pt}
\tablecaption{ Variability between \textit{Spitzer IRS} and MIRI MRS }
\tablehead{
\colhead{Species} & \colhead{Wavelength range ($\mu$m)} & \colhead{Scaling factor}}
\startdata  
Hot H$_2$O & [16.05 -- 16.6], [17.0 -- 17.3] & $2.0^{+0.5}_{-0.5}$ \\
OH & [16.75 -- 16.9] & $2.1^{+1.1}_{-1.1}$ \\
Cold H$_2$O & [21.6 -- 21.9], [23.8 -- 24.0] &  $3.0^{+0.5}_{-0.4}$\\
CO$_2$ & [14.8 -- 15.0] & $3.5^{+0.9}_{-0.9}$ \\
HCN & [13.8 -- 14.1] & $1.0^{+0.5}_{-0.0}$ \\
\enddata
\tablecomments{Scaling factors necessary to match the emission from each species in the best fit model to MIRI MRS, with the \textit{Spitzer IRS} data. The errors correspond to $3\sigma$ confidence intervals from $\chi^2$ curves.}
\end{deluxetable}

\begin{figure*}
    \centering
    \includegraphics[width=0.9\textwidth]{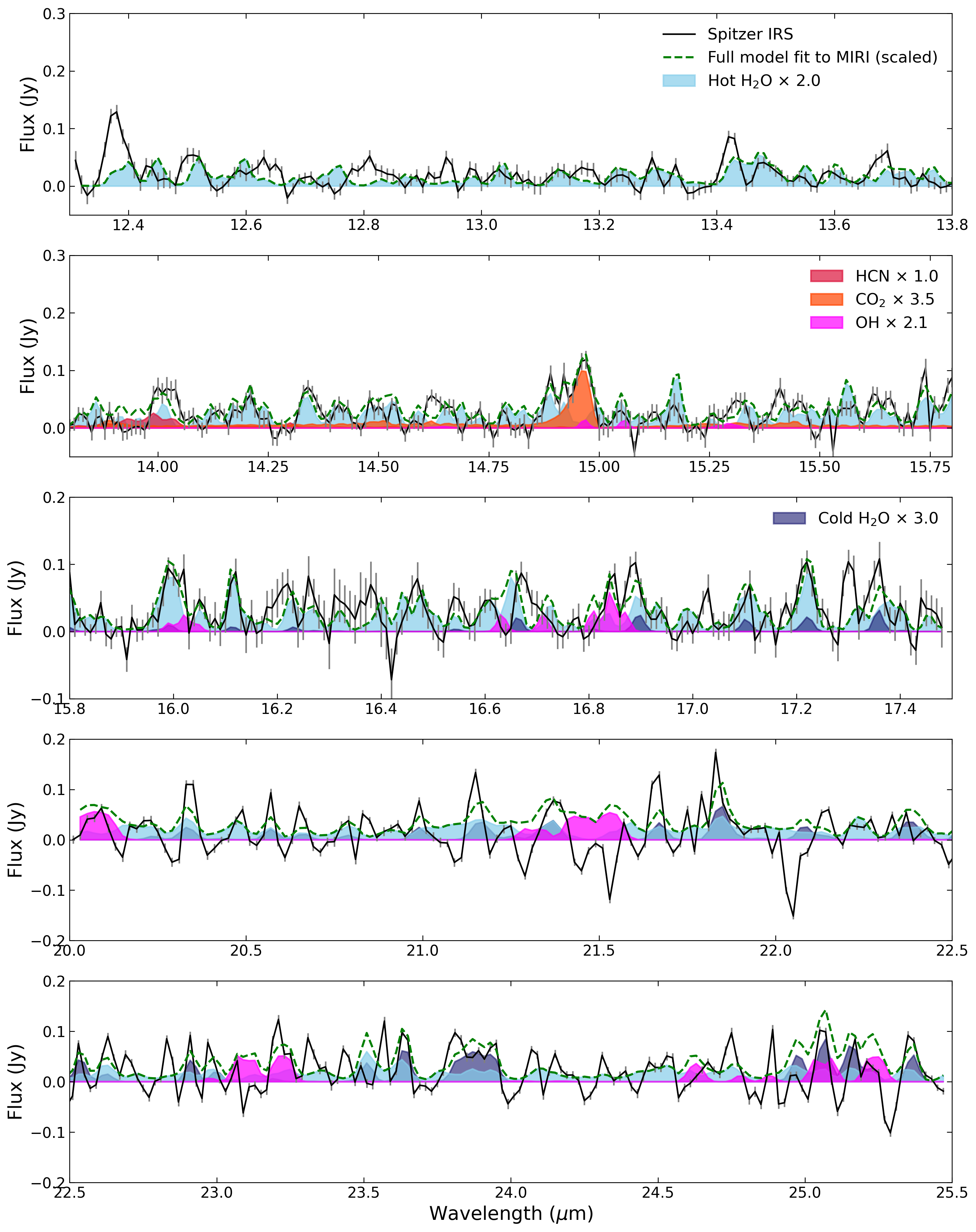}
    \caption{The \textit{Spitzer IRS} spectrum of AS 209 (black line), compared to the best fit model to the MIRI MRS data after scaling the emitting area of each species by a constant factor. The individual contributions from each species are shown as shaded regions.}
    \label{fig:miri_v_mrs_models}
\end{figure*}

The data suggest that the strength of the line emission in the disk of AS 209 has decreased from the epoch of the \textit{Spitzer IRS} observations. Ideally, we would model the old observations to determine the extent of the variability, and whether the emission from some species has varied more than others. However, we are unable to obtain good fits to the \textit{Spitzer IRS} data due to its low resolution and contrast. Rather, we provide an estimate of the variability of each species using the best fit models to the MIRI MRS data.

We estimate the (emitting area) scaling factor necessary to minimize the $\chi^2$ between the downsampled and convolved best fit hot H$_2$O model to MIRI MRS and the \textit{Spitzer IRS} data, using the same wavelength regions as in Section \ref{sec:incontext}. Then, we subtract the scaled hot H$_2$O model from the \textit{Spitzer IRS} data, and find whether we need an additional scaling factor to match the emission from cold H$_2$O, OH, CO$_2$, and HCN. Table \ref{tab:factors} lists the total scaling factors necessary to match the emission from each species, and the wavelength ranges used to estimate each. The latter are chosen based on where the models best reproduce the MIRI MRS data, while trying to avoid regions where lines of different species are strongly blended. We find that the emitting areas of hot H$_2$O and OH have decreased by a factor of 2, while those of cold H$_2$O and CO$_2$ decreased by factors of 3 and 3.5, respectively. On the other hand, HCN does not appear to have changed. Figure \ref{fig:miri_v_mrs_models} shows the \textit{Spitzer IRS} data and the best fit model to MIRI MRS after applying the transformations described above.

It is unclear what physical mechanism could lead to the estimated emission variability. An accretion outburst during the \textit{Spitzer IRS} observations could explain the brighter OH lines, as it would increase the rate of water photo-dissociation. It would also increase the disk temperature, explaining the more extended emitting areas. While there are no accurate accretion luminosity measurements at the time of the \textit{Spitzer IRS} observations, based on the correlation derived by \citet{Banzatti_2023b}, a $3\times$ increase in $L_{acc}$ would be necessary to explain the $2\times$ higher hot H$_2$O line luminosities in the \textit{Spitzer IRS} observations. However, a warmer disk would also produce a brighter continuum at the \textit{Spitzer IRS} epoch, yet we do not see significant differences in the dust emission. At most, the continuum in the \textit{Spitzer IRS} spectrum is 10$\%$ brighter than in the MIRI spectrum, and in some regions they are fully consistent. For comparison, the continuum emission of the well-studied EX Lup disk increased by a factor of $\sim$4 during an outburst which increased $L_{acc}$ by 4.5 \citep{Banzatti_2012}.

An alternative scenario could involve a `leak' or sudden release of a large amount of icy material from an inner disk dust trap at the time of the \textit{Spitzer IRS} observations \citep[e.g.][]{Stammler_2023}. The sublimation of O-rich ices could then explain why the line emission from cold H$_2$O and O-carriers were preferentially affected. Otherwise, the sudden release of volatiles may be related to a giant impact, perhaps due to ongoing planet formation in the inner disk, but this scenario is highly speculative.

Ultimately, a thorough exploration of these and other ideas will be necessary to explain the observed variability, and follow-up monitoring of AS~209 with MIRI MRS may be necessary to the accomplish this goal. Furthermore, we recommend that similar comparisons with \textit{Spitzer IRS} should be carried out as other sources are observed with MIRI MRS \citep[e.g.][]{Schwarz_2023}, which can help determine if this kind of line emission variability is a common phenomenon.

\section{Summary}

We present new observations of the AS~209 disk using JWST MIRI MRS as part of a mid-IR follow-up to the ALMA MAPS Large Program. Our main findings are as follows:

1. The mid-IR spectrum of the AS 209 disk exhibits bright emission from hot (700 - 900 K) water vapor and OH. CO$_2$, HCN, and a colder water vapor component are marginally detected, whereas C$_2$H$_2$ is not detected. We emphasize that the spectrum is severely affected by noise artifacts that prevent an accurate characterization of the molecular gas.

2. Compared to high quality MIRI data of the T-Tauri disks CI Tau and GK Tau, the hot water vapor and OH emission spectra in AS 209 do not have atypical properties. HCN and CO$_2$ emission features are not particularly weaker in the AS 209 spectrum. By contrast, C$_2$H$_2$ line emission in AS 209 is weaker.

3. A comparison with \textit{Spitzer IRS} observations reveals molecular emission variability in the inner disk of AS~209. The emission from hot water vapor and OH decreased by a factor of 2, cold water vapor by a factor of 3, while CO$_2$ emission decreased by a factor of 3.5. HCN does not appear to be affected. We consider an accretion outburst to be unlikely, given the large change in line fluxes but negligible variation in continuum emission. An explanation for the observed variability remains undetermined, but could be related to a triggered sublimation event and sudden release of water and organics during the epoch of the Spitzer observations.

\section*{Acknowledgments}

The authors appreciate the anonymous referee for valuable comments that significantly improved the content, interpretation, and presentation of this work. 

K.I.Ö. acknowledges support from the Simons Foundation (SCOL \#321183), an award from the Simons Foundation (\#321183FY19), and an NSF AAG Grant (\#1907653).

Support for J. H. was provided by NASA through the NASA Hubble Fellowship grant \#HST-HF2-51460.001-A awarded by the Space Telescope Science Institute, which is operated by the Association of Universities for Research in Astronomy, Inc., for NASA, under contract NAS5-26555. 

Support for F. L. was provided by NASA through the NASA Hubble Fellowship grant \#HST-HF2-51512.001-A awarded by the Space Telescope Science Institute, which is operated by the Association of Universities for Research in Astronomy, Incorporated, under NASA contract NAS5-26555. 

Support for C. J. L. was provided by NASA through the NASA Hubble Fellowship grant No. HST-HF2-51535.001-A awarded by the Space Telescope Science Institute, which is operated by the Association of Universities for Research in Astronomy, Inc., for NASA, under contract NAS5-26555.

R. LG. acknowledges support from the Programme National “Physique et Chimie du Milieu Interstellaire” (PCMI) of CNRS/INSU with INC/INP co-funded by CEA and CNES. 

F. M. has received funding from the European Research Council (ERC) under the European Union's Horizon 2020 research and innovation program (grant agreement No. 101053020, project Dust2Planets). 

V.V.G. gratefully acknowledges support from FONDECYT Regular 1221352, ANID BASAL project FB210003, and ANID, -- Millennium Science Initiative Program -- NCN19\_171.

C.W.~acknowledges financial support from the Science and Technology Facilities Council and UK Research and Innovation (grant numbers ST/X001016/1 and MR/T040726/1).

This work is based on observations made with the
NASA/ ESA/CSA James Webb Space Telescope. The data were obtained from the Mikulski Archive for Space Telescopes at the Space Telescope Science Institute, which is operated by the Association of Universities for Research in Astronomy, Incorporated, under NASA contract NAS5-03127. The specific observations analyzed can be accessed via the following \dataset[DOI]{https://doi.org/10.17909/7snz-r967}. Support for Program number (JWST-GO-02025.001-A) was provided through a grant from the STScI under NASA contract NAS5- 03127.

%

\vspace{5mm}
\facilities{JWST (MIRI MRS), Spitzer (IRS)}


\software{\texttt{scipy} \citep{Gommers_2023}, \texttt{astropy} \citep{Robitaille_2021}, \texttt{numpy} \citep{Harris_2020}, \texttt{tinyGP} \citep{Foreman-Mackey_2023}, \texttt{iris} \citep{MirzaRomero_2023}, \texttt{dynesty} \citep{Speagle_2020}
          }



\appendix

\section{Continuum Subtraction}
\label{ap:continuum}
Even after the removal of strong high-frequency fringes using asteroid spectra, we find it unfeasible to visually differentiate true line emission from artifacts, and thus to subtract the continuum emission in the AS~209 spectrum using splines as is customary. We thus develop an iterative Gaussian Processes (GP) continuum-subtraction technique, which does not require the prior manual identification of line-free regions. The procedure is described below. 

First, we mask out negative intensity spikes in the spectrum. To do this we use the peak finding algorithm included in \texttt{scipy} (\texttt{scipy.signal.find\_peaks}). Based on visual inspection of the results, we find good performance by setting a prominence parameter of 0.05 and distance between peaks of 2 pixels\footnote{For a detailed description of the algorithm and meaning of each parameter see: docs.scipy.org/doc/scipy/reference} based on visual inspection of the data. Second, we condition a Gaussian Process to the masked data. In particular, we use a linear combination of three squared exponential covariance functions (i.e. kernels) $k_1, k_2, k_3$, such that each $k_n$ is defined as:
\begin{equation}
    k_n = \alpha_n \exp{\left( -\frac{(x_i-x_j)^2}{2 \lambda_n^2} \right)},
\end{equation}
where $\alpha$ and $\lambda$ are the amplitude and correlation length scale, respectively, and $(x_i, x_j)$ represent any pair of points in the wavelength grid. With $k_1$, we aim to capture the overall shape of the flux baseline, and we set $\alpha_1 = 100$ mJy and $\lambda_1 = 1.0$ $\mu$m. Then, $k_2$ is set to model  dust features and artifacts with $\alpha_2 = 10$ mJy, but we constrain the length scale $\lambda_2 = 0.5$ $\mu$m to prevent it from erroneously removing broad emission features produced by the overlap of lines. $k_3$ is finally used to model the molecular line emission by setting  $\alpha_3 = 1$ mJy and $\lambda_3 = 0.001$ $\mu$m. These values are chosen based on prior predictive checks and visual inspection of the decomposition result. We experiment using different kernels like Matern covariance functions; the resulting decomposition is not overly sensitive to the exact choice of kernels as long as we use a mixture of functions with long and short correlation length scales, and ultimately we choose the exponential squared for its simplicity. 

\begin{figure*}
    \centering
    \includegraphics[width=0.9\textwidth]{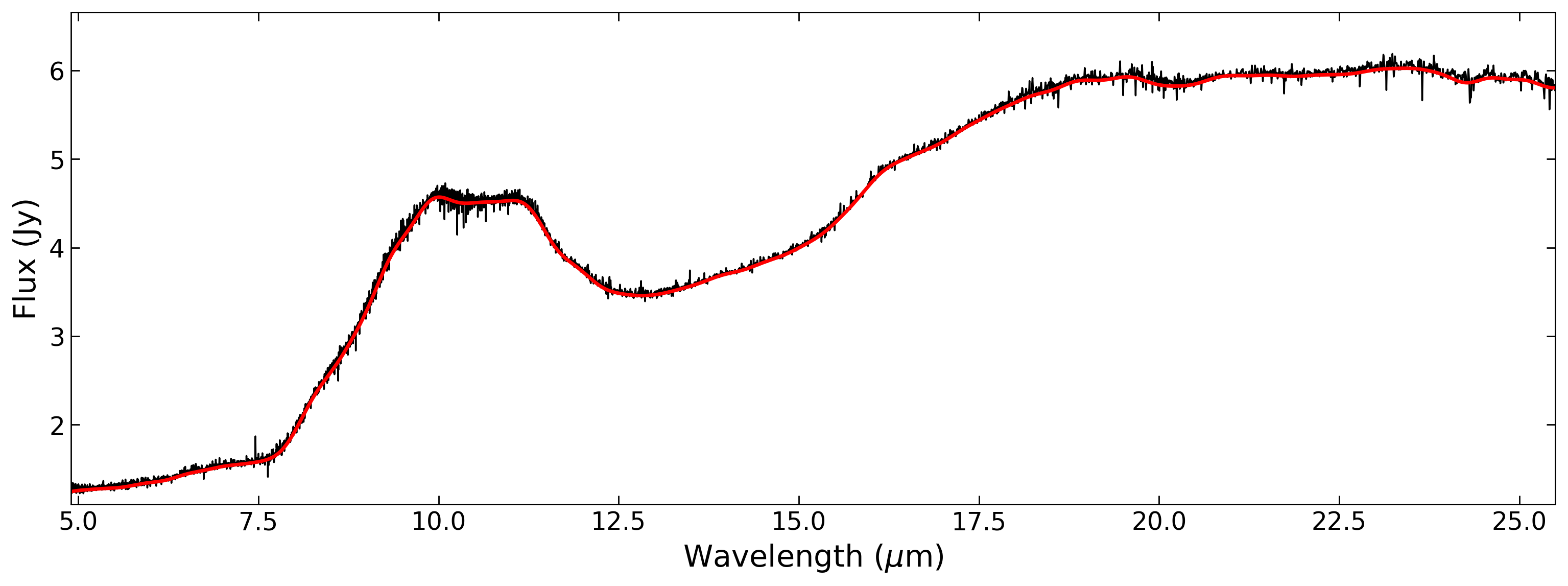}
    \caption{The MIRI MRS spectrum of AS~209 in black, and the contributions from the conditioned $k_1 + k_2$ squared exponential kernels (i.e. the continuum baseline) in red.}
    \label{fig:continuum}
\end{figure*}

\begin{figure*}
    \centering
    \includegraphics[width=0.9\textwidth]{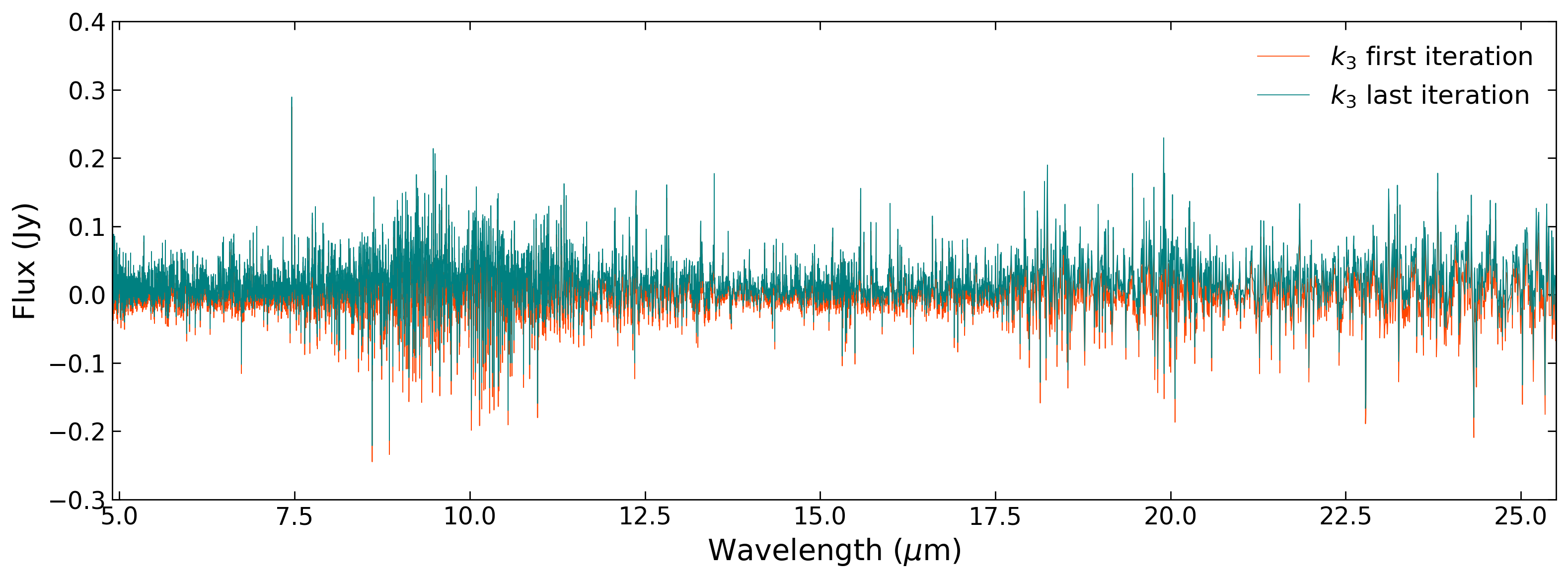}
    \caption{The contributions from the $k_3$ kernel after the GP model has been conditioned to the data. The model in orange shows the first iteration, and the model in teal shows the last iteration. Note that the conditioned $k_3$ model is shifted towards positive values.}
    \label{fig:k1}
\end{figure*}

The model is conditioned on the data using the GP Python package \texttt{tinyGP} \citep{Foreman-Mackey_2023} assuming, initially, a mean of zero. By setting the mean of the process to zero, the conditioned continuum function described by $k_1 + k_2$ does not trace a baseline under the line emission, but rather it follows the mean of the spectrum. Consequently, the line emission modeled by $k_3$ is incorrectly centered around zero. To solve this, we set all points of the conditioned $k_3$ contributions with negative values to zero, and repeat the GP conditioning using the contributions from $k_3$ as the new mean function of the process. At each iteration, the location of the continuum baseline is thus shifted downward by a small amount, but since $\lambda_1$ and $\lambda_2$ are constrained, the baseline is prevented from over-fitting the data and shifting the entire continuum-subtracted spectrum above zero. The process is repeated until the change in the standard deviation of the negative part of the conditioned $k_3$ model falls below 0.5 mJy (5 iterations). Finally, the resulting continuum baseline is linearly interpolated back onto the full wavelength grid and subtracted from the MIRI MRS spectrum, including those pixels flagged in the first step. We emphasize that the conditioned $k_3$ itself is not used as the continuum-subtracted spectrum, but rather the conditioned $k_1 + k_2$ kernels subtracted from the data.


\bibliography{sample631}{}

\begin{thebibliography}{}
\expandafter\ifx\csname natexlab\endcsname\relax\def\natexlab#1{#1}\fi
\providecommand{\url}[1]{\href{#1}{#1}}
\providecommand{\dodoi}[1]{doi:~\href{http://doi.org/#1}{\nolinkurl{#1}}}
\providecommand{\doeprint}[1]{\href{http://ascl.net/#1}{\nolinkurl{http://ascl.net/#1}}}
\providecommand{\doarXiv}[1]{\href{https://arxiv.org/abs/#1}{\nolinkurl{https://arxiv.org/abs/#1}}}

\bibitem[{{Andrews} {et~al.}(2018){Andrews}, {Huang}, {P{\'e}rez}, {Isella}, {Dullemond}, {Kurtovic}, {Guzm{\'a}n}, {Carpenter}, {Wilner}, {Zhang}, {Zhu}, {Birnstiel}, {Bai}, {Benisty}, {Hughes}, {{\"O}berg}, \& {Ricci}}]{Andrews_2018}
{Andrews}, S.~M., {Huang}, J., {P{\'e}rez}, L.~M., {et~al.} 2018, \apjl, 869, L41, \dodoi{10.3847/2041-8213/aaf741}

\bibitem[{{Antonellini} {et~al.}(2015){Antonellini}, {Kamp}, {Riviere-Marichalar}, {Meijerink}, {Woitke}, {Thi}, {Spaans}, {Aresu}, \& {Lee}}]{Antonellini_2015}
{Antonellini}, S., {Kamp}, I., {Riviere-Marichalar}, P., {et~al.} 2015, \aap, 582, A105, \dodoi{10.1051/0004-6361/201525724}

\bibitem[{{Bae} {et~al.}(2022){Bae}, {Teague}, {Andrews}, {Benisty}, {Facchini}, {Galloway-Sprietsma}, {Loomis}, {Aikawa}, {Alarc{\'o}n}, {Bergin}, {Bergner}, {Booth}, {Cataldi}, {Cleeves}, {Czekala}, {Guzm{\'a}n}, {Huang}, {Ilee}, {Kurtovic}, {Law}, {Le Gal}, {Liu}, {Long}, {M{\'e}nard}, {{\"O}berg}, {P{\'e}rez}, {Qi}, {Schwarz}, {Sierra}, {Walsh}, {Wilner}, \& {Zhang}}]{Bae_2022}
{Bae}, J., {Teague}, R., {Andrews}, S.~M., {et~al.} 2022, \apjl, 934, L20, \dodoi{10.3847/2041-8213/ac7fa3}

\bibitem[{{Banzatti} {et~al.}(2014){Banzatti}, {Meyer}, {Manara}, {Pontoppidan}, \& {Testi}}]{Banzatti_2014}
{Banzatti}, A., {Meyer}, M.~R., {Manara}, C.~F., {Pontoppidan}, K.~M., \& {Testi}, L. 2014, \apj, 780, 26, \dodoi{10.1088/0004-637X/780/1/26}

\bibitem[{{Banzatti} {et~al.}(2012){Banzatti}, {Meyer}, {Bruderer}, {Geers}, {Pascucci}, {Lahuis}, {Juh{\'a}sz}, {Henning}, \& {{\'A}brah{\'a}m}}]{Banzatti_2012}
{Banzatti}, A., {Meyer}, M.~R., {Bruderer}, S., {et~al.} 2012, \apj, 745, 90, \dodoi{10.1088/0004-637X/745/1/90}

\bibitem[{{Banzatti} {et~al.}(2020){Banzatti}, {Pascucci}, {Bosman}, {Pinilla}, {Salyk}, {Herczeg}, {Pontoppidan}, {Vazquez}, {Watkins}, {Krijt}, {Hendler}, \& {Long}}]{Banzatti_2020}
{Banzatti}, A., {Pascucci}, I., {Bosman}, A.~D., {et~al.} 2020, \apj, 903, 124, \dodoi{10.3847/1538-4357/abbc1a}

\bibitem[{{Banzatti} {et~al.}(2023{\natexlab{a}}){Banzatti}, {Pontoppidan}, {P{\'e}re Ch{\'a}vez}, {Salyk}, {Diehl}, {Bruderer}, {Herczeg}, {Carmona}, {Pascucci}, {Brittain}, {Jensen}, {Grant}, {van Dishoeck}, {Kamp}, {Bosman}, {{\"O}berg}, {Blake}, {Meyer}, {Gaidos}, {Boogert}, {Rayner}, \& {Wheeler}}]{Banzatti_2023}
{Banzatti}, A., {Pontoppidan}, K.~M., {P{\'e}re Ch{\'a}vez}, J., {et~al.} 2023{\natexlab{a}}, \aj, 165, 72, \dodoi{10.3847/1538-3881/aca80b}

\bibitem[{{Banzatti} {et~al.}(2023{\natexlab{b}}){Banzatti}, {Pontoppidan}, {Carr}, {Jellison}, {Pascucci}, {Najita}, {Mirza-Romero}, {Oberg}, {Kalyaan}, {Pinilla}, {Krijt}, {Long}, {Lambrechts}, {Rosotti}, {Herczeg}, {Salyk}, {Zhang}, {Ballering}, {Meyer}, {Bruderer}, \& {the JDISCS collaboration}}]{Banzatti_2023b}
{Banzatti}, A., {Pontoppidan}, K.~M., {Carr}, J., {et~al.} 2023{\natexlab{b}}, arXiv e-prints, arXiv:2307.03846, \dodoi{10.48550/arXiv.2307.03846}

\bibitem[{{Bergner} {et~al.}(2018){Bergner}, {Guzm{\'a}n}, {{\"O}berg}, {Loomis}, \& {Pegues}}]{Bergner_2018}
{Bergner}, J.~B., {Guzm{\'a}n}, V.~G., {{\"O}berg}, K.~I., {Loomis}, R.~A., \& {Pegues}, J. 2018, \apj, 857, 69, \dodoi{10.3847/1538-4357/aab664}

\bibitem[{{Bern{\'e}} {et~al.}(2023){Bern{\'e}}, {Martin-Drumel}, {Schroetter}, {Goicoechea}, {Jacovella}, {Gans}, {Dartois}, {Coudert}, {Bergin}, {Alarcon}, {Cami}, {Roueff}, {Black}, {Asvany}, {Habart}, {Peeters}, {Canin}, {Trahin}, {Joblin}, {Schlemmer}, {Thorwirth}, {Cernicharo}, {Gerin}, {Tielens}, {Zannese}, {Abergel}, {Bernard-Salas}, {Boersma}, {Bron}, {Chown}, {Cuadrado}, {Dicken}, {Elyajouri}, {Fuente}, {Gordon}, {Issa}, {Kannavou}, {Khan}, {Lacinbala}, {Languignon}, {Le Gal}, {Maragkoudakis}, {Meshaka}, {Okada}, {Onaka}, {Pasquini}, {Pound}, {Robberto}, {R{\"o}llig}, {Schefter}, {Schirmer}, {Sidhu}, {Tabone}, {Van De Putte}, {Vicente}, \& {Wolfire}}]{berne_2023}
{Bern{\'e}}, O., {Martin-Drumel}, M.-A., {Schroetter}, I., {et~al.} 2023, \nat, 621, 56, \dodoi{10.1038/s41586-023-06307-x}

\bibitem[{{Birnstiel} {et~al.}(2010){Birnstiel}, {Ricci}, {Trotta}, {Dullemond}, {Natta}, {Testi}, {Dominik}, {Henning}, {Ormel}, \& {Zsom}}]{Birnstiel_2010}
{Birnstiel}, T., {Ricci}, L., {Trotta}, F., {et~al.} 2010, \aap, 516, L14, \dodoi{10.1051/0004-6361/201014893}

\bibitem[{{Blevins} {et~al.}(2016){Blevins}, {Pontoppidan}, {Banzatti}, {Zhang}, {Najita}, {Carr}, {Salyk}, \& {Blake}}]{Blevins_2016}
{Blevins}, S.~M., {Pontoppidan}, K.~M., {Banzatti}, A., {et~al.} 2016, \apj, 818, 22, \dodoi{10.3847/0004-637X/818/1/22}

\bibitem[{{Bosman} {et~al.}(2022){Bosman}, {Bergin}, {Calahan}, \& {Duval}}]{Bosman_2022}
{Bosman}, A.~D., {Bergin}, E.~A., {Calahan}, J., \& {Duval}, S.~E. 2022, \apjl, 930, L26, \dodoi{10.3847/2041-8213/ac66ce}

\bibitem[{Bushouse {et~al.}(2023)Bushouse, Eisenhamer, Dencheva, Davies, Greenfield, Morrison, Hodge, Simon, Grumm, Droettboom, Slavich, Sosey, Pauly, Miller, Jedrzejewski, Hack, Davis, Crawford, Law, Gordon, Regan, Cara, MacDonald, Bradley, Shanahan, Jamieson, Teodoro, \& Williams}]{Bushouse_2023}
Bushouse, H., Eisenhamer, J., Dencheva, N., {et~al.} 2023, JWST Calibration Pipeline, 1.11.0,  Zenodo, \dodoi{10.5281/zenodo.8067394}

\bibitem[{{Carnall}(2021)}]{Carnall_2021}
{Carnall}, A. 2021, {SpectRes: Simple spectral resampling}, Astrophysics Source Code Library, record ascl:2104.019.
\newblock \doeprint{2104.019}

\bibitem[{{Carr} \& {Najita}(2008)}]{Carr_2008}
{Carr}, J.~S., \& {Najita}, J.~R. 2008, Science, 319, 1504, \dodoi{10.1126/science.1153807}

\bibitem[{{Carr} \& {Najita}(2011)}]{Carr_2011}
---. 2011, \apj, 733, 102, \dodoi{10.1088/0004-637X/733/2/102}

\bibitem[{{Carr} \& {Najita}(2014)}]{Carr_2014}
---. 2014, \apj, 788, 66, \dodoi{10.1088/0004-637X/788/1/66}

\bibitem[{{Carr} {et~al.}(2004){Carr}, {Tokunaga}, \& {Najita}}]{Carr_2004}
{Carr}, J.~S., {Tokunaga}, A.~T., \& {Najita}, J. 2004, \apj, 603, 213, \dodoi{10.1086/381356}

\bibitem[{{Ciesla} \& {Cuzzi}(2006)}]{Ciesla_2006}
{Ciesla}, F.~J., \& {Cuzzi}, J.~N. 2006, \icarus, 181, 178, \dodoi{10.1016/j.icarus.2005.11.009}

\bibitem[{{Dominik} \& {Tielens}(1997)}]{Dominik_1997}
{Dominik}, C., \& {Tielens}, A.~G.~G.~M. 1997, \apj, 480, 647, \dodoi{10.1086/303996}

\bibitem[{{Du} {et~al.}(2017){Du}, {Bergin}, {Hogerheijde}, {van Dishoeck}, {Blake}, {Bruderer}, {Cleeves}, {Dominik}, {Fedele}, {Lis}, {Melnick}, {Neufeld}, {Pearson}, \& {Y{\i}ld{\i}z}}]{Du_2017}
{Du}, F., {Bergin}, E.~A., {Hogerheijde}, M., {et~al.} 2017, \apj, 842, 98, \dodoi{10.3847/1538-4357/aa70ee}

\bibitem[{{Evans} {et~al.}(2003){Evans}, {Allen}, {Blake}, {Boogert}, {Bourke}, {Harvey}, {Kessler}, {Koerner}, {Lee}, {Mundy}, {Myers}, {Padgett}, {Pontoppidan}, {Sargent}, {Stapelfeldt}, {van Dishoeck}, {Young}, \& {Young}}]{Evans_2003}
{Evans}, Neal~J., I., {Allen}, L.~E., {Blake}, G.~A., {et~al.} 2003, \pasp, 115, 965, \dodoi{10.1086/376697}

\bibitem[{{Fang} {et~al.}(2018){Fang}, {Pascucci}, {Edwards}, {Gorti}, {Banzatti}, {Flock}, {Hartigan}, {Herczeg}, \& {Dupree}}]{Fang_2018}
{Fang}, M., {Pascucci}, I., {Edwards}, S., {et~al.} 2018, \apj, 868, 28, \dodoi{10.3847/1538-4357/aae780}

\bibitem[{{Favre} {et~al.}(2019){Favre}, {Fedele}, {Maud}, {Booth}, {Tazzari}, {Miotello}, {Testi}, {Semenov}, \& {Bruderer}}]{Favre_2019}
{Favre}, C., {Fedele}, D., {Maud}, L., {et~al.} 2019, \apj, 871, 107, \dodoi{10.3847/1538-4357/aaf80c}

\bibitem[{{Fedele} {et~al.}(2011){Fedele}, {Pascucci}, {Brittain}, {Kamp}, {Woitke}, {Williams}, {Dent}, \& {Thi}}]{Fedele_2011}
{Fedele}, D., {Pascucci}, I., {Brittain}, S., {et~al.} 2011, \apj, 732, 106, \dodoi{10.1088/0004-637X/732/2/106}

\bibitem[{{Foreman-Mackey} {et~al.}(2022){Foreman-Mackey}, {Yadav}, {Theorashid}, {Fowlie}, {Tronsgaard}, {Schmerler}, \& {Killestein}}]{Foreman-Mackey_2023}
{Foreman-Mackey}, D., {Yadav}, S., {Theorashid}, {et~al.} 2022, {dfm/tinygp: v0.2.3}, v0.2.3, Zenodo,  Zenodo, \dodoi{10.5281/zenodo.7269074}

\bibitem[{{Gaia Collaboration} {et~al.}(2018){Gaia Collaboration}, {Brown}, {Vallenari}, {Prusti}, {de Bruijne}, {Babusiaux}, {Bailer-Jones}, {Biermann}, {Evans}, {Eyer}, {Jansen}, {Jordi}, {Klioner}, {Lammers}, {Lindegren}, {Luri}, {Mignard}, {Panem}, {Pourbaix}, {Randich}, {Sartoretti}, {Siddiqui}, {Soubiran}, {van Leeuwen}, {Walton}, {Arenou}, {Bastian}, {Cropper}, {Drimmel}, {Katz}, {Lattanzi}, {Bakker}, {Cacciari}, {Casta{\~n}eda}, {Chaoul}, {Cheek}, {De Angeli}, {Fabricius}, {Guerra}, {Holl}, {Masana}, {Messineo}, {Mowlavi}, {Nienartowicz}, {Panuzzo}, {Portell}, {Riello}, {Seabroke}, {Tanga}, {Th{\'e}venin}, {Gracia-Abril}, {Comoretto}, {Garcia-Reinaldos}, {Teyssier}, {Altmann}, {Andrae}, {Audard}, {Bellas-Velidis}, {Benson}, {Berthier}, {Blomme}, {Burgess}, {Busso}, {Carry}, {Cellino}, {Clementini}, {Clotet}, {Creevey}, {Davidson}, {De Ridder}, {Delchambre}, {Dell'Oro}, {Ducourant}, {Fern{\'a}ndez-Hern{\'a}ndez}, {Fouesneau}, {Fr{\'e}mat}, {Galluccio}, {Garc{\'\i}a-Torres},
  {Gonz{\'a}lez-N{\'u}{\~n}ez}, {Gonz{\'a}lez-Vidal}, {Gosset}, {Guy}, {Halbwachs}, {Hambly}, {Harrison}, {Hern{\'a}ndez}, {Hestroffer}, {Hodgkin}, {Hutton}, {Jasniewicz}, {Jean-Antoine-Piccolo}, {Jordan}, {Korn}, {Krone-Martins}, {Lanzafame}, {Lebzelter}, {L{\"o}ffler}, {Manteiga}, {Marrese}, {Mart{\'\i}n-Fleitas}, {Moitinho}, {Mora}, {Muinonen}, {Osinde}, {Pancino}, {Pauwels}, {Petit}, {Recio-Blanco}, {Richards}, {Rimoldini}, {Robin}, {Sarro}, {Siopis}, {Smith}, {Sozzetti}, {S{\"u}veges}, {Torra}, {van Reeven}, {Abbas}, {Abreu Aramburu}, {Accart}, {Aerts}, {Altavilla}, {{\'A}lvarez}, {Alvarez}, {Alves}, {Anderson}, {Andrei}, {Anglada Varela}, {Antiche}, {Antoja}, {Arcay}, {Astraatmadja}, {Bach}, {Baker}, {Balaguer-N{\'u}{\~n}ez}, {Balm}, {Barache}, {Barata}, {Barbato}, {Barblan}, {Barklem}, {Barrado}, {Barros}, {Barstow}, {Bartholom{\'e} Mu{\~n}oz}, {Bassilana}, {Becciani}, {Bellazzini}, {Berihuete}, {Bertone}, {Bianchi}, {Bienaym{\'e}}, {Blanco-Cuaresma}, {Boch}, {Boeche}, {Bombrun}, {Borrachero},
  {Bossini}, {Bouquillon}, {Bourda}, {Bragaglia}, {Bramante}, {Breddels}, {Bressan}, {Brouillet}, {Br{\"u}semeister}, {Brugaletta}, {Bucciarelli}, {Burlacu}, {Busonero}, {Butkevich}, {Buzzi}, {Caffau}, {Cancelliere}, {Cannizzaro}, {Cantat-Gaudin}, {Carballo}, {Carlucci}, {Carrasco}, {Casamiquela}, {Castellani}, {Castro-Ginard}, {Charlot}, {Chemin}, {Chiavassa}, {Cocozza}, {Costigan}, {Cowell}, {Crifo}, {Crosta}, {Crowley}, {Cuypers}, {Dafonte}, {Damerdji}, {Dapergolas}, {David}, {David}, {de Laverny}, {De Luise}, {De March}, {de Martino}, {de Souza}, {de Torres}, {Debosscher}, {del Pozo}, {Delbo}, {Delgado}, {Delgado}, {Di Matteo}, {Diakite}, {Diener}, {Distefano}, {Dolding}, {Drazinos}, {Dur{\'a}n}, {Edvardsson}, {Enke}, {Eriksson}, {Esquej}, {Eynard Bontemps}, {Fabre}, {Fabrizio}, {Faigler}, {Falc{\~a}o}, {Farr{\`a}s Casas}, {Federici}, {Fedorets}, {Fernique}, {Figueras}, {Filippi}, {Findeisen}, {Fonti}, {Fraile}, {Fraser}, {Fr{\'e}zouls}, {Gai}, {Galleti}, {Garabato}, {Garc{\'\i}a-Sedano}, {Garofalo},
  {Garralda}, {Gavel}, {Gavras}, {Gerssen}, {Geyer}, {Giacobbe}, {Gilmore}, {Girona}, {Giuffrida}, {Glass}, {Gomes}, {Granvik}, {Gueguen}, {Guerrier}, {Guiraud}, {Guti{\'e}rrez-S{\'a}nchez}, {Haigron}, {Hatzidimitriou}, {Hauser}, {Haywood}, {Heiter}, {Helmi}, {Heu}, {Hilger}, {Hobbs}, {Hofmann}, {Holland}, {Huckle}, {Hypki}, {Icardi}, {Jan{\ss}en}, {Jevardat de Fombelle}, {Jonker}, {Juh{\'a}sz}, {Julbe}, {Karampelas}, {Kewley}, {Klar}, {Kochoska}, {Kohley}, {Kolenberg}, {Kontizas}, {Kontizas}, {Koposov}, {Kordopatis}, {Kostrzewa-Rutkowska}, {Koubsky}, {Lambert}, {Lanza}, {Lasne}, {Lavigne}, {Le Fustec}, {Le Poncin-Lafitte}, {Lebreton}, {Leccia}, {Leclerc}, {Lecoeur-Taibi}, {Lenhardt}, {Leroux}, {Liao}, {Licata}, {Lindstr{\o}m}, {Lister}, {Livanou}, {Lobel}, {L{\'o}pez}, {Managau}, {Mann}, {Mantelet}, {Marchal}, {Marchant}, {Marconi}, {Marinoni}, {Marschalk{\'o}}, {Marshall}, {Martino}, {Marton}, {Mary}, {Massari}, {Matijevi{\v{c}}}, {Mazeh}, {McMillan}, {Messina}, {Michalik}, {Millar}, {Molina}, {Molinaro},
  {Moln{\'a}r}, {Montegriffo}, {Mor}, {Morbidelli}, {Morel}, {Morris}, {Mulone}, {Muraveva}, {Musella}, {Nelemans}, {Nicastro}, {Noval}, {O'Mullane}, {Ord{\'e}novic}, {Ord{\'o}{\~n}ez-Blanco}, {Osborne}, {Pagani}, {Pagano}, {Pailler}, {Palacin}, {Palaversa}, {Panahi}, {Pawlak}, {Piersimoni}, {Pineau}, {Plachy}, {Plum}, {Poggio}, {Poujoulet}, {Pr{\v{s}}a}, {Pulone}, {Racero}, {Ragaini}, {Rambaux}, {Ramos-Lerate}, {Regibo}, {Reyl{\'e}}, {Riclet}, {Ripepi}, {Riva}, {Rivard}, {Rixon}, {Roegiers}, {Roelens}, {Romero-G{\'o}mez}, {Rowell}, {Royer}, {Ruiz-Dern}, {Sadowski}, {Sagrist{\`a} Sell{\'e}s}, {Sahlmann}, {Salgado}, {Salguero}, {Sanna}, {Santana-Ros}, {Sarasso}, {Savietto}, {Schultheis}, {Sciacca}, {Segol}, {Segovia}, {S{\'e}gransan}, {Shih}, {Siltala}, {Silva}, {Smart}, {Smith}, {Solano}, {Solitro}, {Sordo}, {Soria Nieto}, {Souchay}, {Spagna}, {Spoto}, {Stampa}, {Steele}, {Steidelm{\"u}ller}, {Stephenson}, {Stoev}, {Suess}, {Surdej}, {Szabados}, {Szegedi-Elek}, {Tapiador}, {Taris}, {Tauran}, {Taylor},
  {Teixeira}, {Terrett}, {Teyssandier}, {Thuillot}, {Titarenko}, {Torra Clotet}, {Turon}, {Ulla}, {Utrilla}, {Uzzi}, {Vaillant}, {Valentini}, {Valette}, {van Elteren}, {Van Hemelryck}, {van Leeuwen}, {Vaschetto}, {Vecchiato}, {Veljanoski}, {Viala}, {Vicente}, {Vogt}, {von Essen}, {Voss}, {Votruba}, {Voutsinas}, {Walmsley}, {Weiler}, {Wertz}, {Wevers}, {Wyrzykowski}, {Yoldas}, {{\v{Z}}erjal}, {Ziaeepour}, {Zorec}, {Zschocke}, {Zucker}, {Zurbach}, \& {Zwitter}}]{Gaia}
{Gaia Collaboration}, {Brown}, A.~G.~A., {Vallenari}, A., {et~al.} 2018, \aap, 616, A1, \dodoi{10.1051/0004-6361/201833051}

\bibitem[{{Gommers} {et~al.}(2023){Gommers}, {Virtanen}, {Burovski}, {Haberland}, {Weckesser}, {Oliphant}, {Reddy}, {Cournapeau}, {Alexbrc}, {Nelson}, {Peterson}, {Wilson}, {Roy}, {Endolith}, {Polat}, {Mayorov}, {Van Der Walt}, {Brett}, {Laxalde}, {Larson}, {Millman}, {Sakai}, {Lars}, {Peterbell10}, {Van Mulbregt}, {Carey}, {Eric-Jones}, {McKibben}, {Kai}, \& {Kern}}]{Gommers_2023}
{Gommers}, R., {Virtanen}, P., {Burovski}, E., {et~al.} 2023, {scipy/scipy: SciPy 1.10.1}, v1.10.1, Zenodo,  Zenodo, \dodoi{10.5281/zenodo.7655153}

\bibitem[{{Gordon} {et~al.}(2022){Gordon}, {Rothman}, {Hargreaves}, {Hashemi}, {Karlovets}, {Skinner}, {Conway}, {Hill}, {Kochanov}, {Tan}, {Wcis{\l}o}, {Finenko}, {Nelson}, {Bernath}, {Birk}, {Boudon}, {Campargue}, {Chance}, {Coustenis}, {Drouin}, {Flaud}, {Gamache}, {Hodges}, {Jacquemart}, {Mlawer}, {Nikitin}, {Perevalov}, {Rotger}, {Tennyson}, {Toon}, {Tran}, {Tyuterev}, {Adkins}, {Baker}, {Barbe}, {Can{\`e}}, {Cs{\'a}sz{\'a}r}, {Dudaryonok}, {Egorov}, {Fleisher}, {Fleurbaey}, {Foltynowicz}, {Furtenbacher}, {Harrison}, {Hartmann}, {Horneman}, {Huang}, {Karman}, {Karns}, {Kassi}, {Kleiner}, {Kofman}, {Kwabia-Tchana}, {Lavrentieva}, {Lee}, {Long}, {Lukashevskaya}, {Lyulin}, {Makhnev}, {Matt}, {Massie}, {Melosso}, {Mikhailenko}, {Mondelain}, {M{\"u}ller}, {Naumenko}, {Perrin}, {Polyansky}, {Raddaoui}, {Raston}, {Reed}, {Rey}, {Richard}, {T{\'o}bi{\'a}s}, {Sadiek}, {Schwenke}, {Starikova}, {Sung}, {Tamassia}, {Tashkun}, {Vander Auwera}, {Vasilenko}, {Vigasin}, {Villanueva}, {Vispoel}, {Wagner}, {Yachmenev}, \&
  {Yurchenko}}]{Gordon_2022}
{Gordon}, I.~E., {Rothman}, L.~S., {Hargreaves}, R.~J., {et~al.} 2022, \jqsrt, 277, 107949, \dodoi{10.1016/j.jqsrt.2021.107949}

\bibitem[{{Grant} {et~al.}(2022){Grant}, {van Dishoeck}, {Tabone}, {Gasman}, {Henning}, {Kamp}, {G{\"u}del}, {Lagage}, {Bettoni}, {Perotti}, {Christiaens}, {Samland}, {Arabhavi}, {Argyriou}, {Abergel}, {Absil}, {Barrado}, {Boccaletti}, {Bouwman}, {Garatti}, {Geers}, {Glauser}, {Guadarrama}, {Jang}, {Kanwar}, {Lahuis}, {Morales-Calder{\'o}n}, {Mueller}, {Nehm{\'e}}, {Olofsson}, {Pantin}, {Pawellek}, {Ray}, {Rodgers-Lee}, {Scheithauer}, {Schreiber}, {Schwarz}, {Temmink}, {Vandenbussche}, {Vlasblom}, {Waters}, {Wright}, {Colina}, {Greve}, {Justannont}, \& {{\"O}stlin}}]{Grant_2022}
{Grant}, S.~L., {van Dishoeck}, E.~F., {Tabone}, B., {et~al.} 2022, arXiv e-prints, arXiv:2212.08047, \dodoi{10.48550/arXiv.2212.08047}

\bibitem[{{Gundlach} \& {Blum}(2015)}]{Gundlach_2015}
{Gundlach}, B., \& {Blum}, J. 2015, \apj, 798, 34, \dodoi{10.1088/0004-637X/798/1/34}

\bibitem[{{Guzm{\'a}n} {et~al.}(2018){Guzm{\'a}n}, {Huang}, {Andrews}, {Isella}, {P{\'e}rez}, {Carpenter}, {Dullemond}, {Ricci}, {Birnstiel}, {Zhang}, {Zhu}, {Bai}, {Benisty}, {{\"O}berg}, \& {Wilner}}]{guzman_2018}
{Guzm{\'a}n}, V.~V., {Huang}, J., {Andrews}, S.~M., {et~al.} 2018, \apjl, 869, L48, \dodoi{10.3847/2041-8213/aaedae}

\bibitem[{{Harris} {et~al.}(2020){Harris}, {Millman}, {van der Walt}, {Gommers}, {Virtanen}, {Cournapeau}, {Wieser}, {Taylor}, {Berg}, {Smith}, {Kern}, {Picus}, {Hoyer}, {van Kerkwijk}, {Brett}, {Haldane}, {del R{\'\i}o}, {Wiebe}, {Peterson}, {G{\'e}rard-Marchant}, {Sheppard}, {Reddy}, {Weckesser}, {Abbasi}, {Gohlke}, \& {Oliphant}}]{Harris_2020}
{Harris}, C.~R., {Millman}, K.~J., {van der Walt}, S.~J., {et~al.} 2020, \nat, 585, 357, \dodoi{10.1038/s41586-020-2649-2}

\bibitem[{{Herbst} \& {van Dishoeck}(2009)}]{Herbst_2009}
{Herbst}, E., \& {van Dishoeck}, E.~F. 2009, \araa, 47, 427, \dodoi{10.1146/annurev-astro-082708-101654}

\bibitem[{{Hogerheijde} {et~al.}(2011){Hogerheijde}, {Bergin}, {Brinch}, {Cleeves}, {Fogel}, {Blake}, {Dominik}, {Lis}, {Melnick}, {Neufeld}, {Pani{\'c}}, {Pearson}, {Kristensen}, {Y{\i}ld{\i}z}, \& {van Dishoeck}}]{Hogerheijde_2011}
{Hogerheijde}, M.~R., {Bergin}, E.~A., {Brinch}, C., {et~al.} 2011, Science, 334, 338, \dodoi{10.1126/science.1208931}

\bibitem[{{Houck} {et~al.}(2004){Houck}, {Roellig}, {Van Cleve}, {Forrest}, {Herter}, {Lawrence}, {Matthews}, {Reitsema}, {Soifer}, {Watson}, {Weedman}, {Huisjen}, {Troeltzsch}, {Barry}, {Bernard-Salas}, {Blacken}, {Brandl}, {Charmandaris}, {Devost}, {Gull}, {Hall}, {Henderson}, {Higdon}, {Pirger}, {Schoenwald}, {Sloan}, {Uchida}, {Appleton}, {Armus}, {Burgdorf}, {Fajardo-Acosta}, {Grillmair}, {Ingalls}, {Morris}, \& {Teplitz}}]{Houck_2004}
{Houck}, J.~R., {Roellig}, T.~L., {Van Cleve}, J., {et~al.} 2004, in Society of Photo-Optical Instrumentation Engineers (SPIE) Conference Series, Vol. 5487, Optical, Infrared, and Millimeter Space Telescopes, ed. J.~C. {Mather}, 62--76, \dodoi{10.1117/12.550517}

\bibitem[{{Huang} {et~al.}(2017){Huang}, {{\"O}berg}, {Qi}, {Aikawa}, {Andrews}, {Furuya}, {Guzm{\'a}n}, {Loomis}, {van Dishoeck}, \& {Wilner}}]{Huang_2017}
{Huang}, J., {{\"O}berg}, K.~I., {Qi}, C., {et~al.} 2017, \apj, 835, 231, \dodoi{10.3847/1538-4357/835/2/231}

\bibitem[{{Huang} {et~al.}(2018){Huang}, {Andrews}, {Dullemond}, {Isella}, {P{\'e}rez}, {Guzm{\'a}n}, {{\"O}berg}, {Zhu}, {Zhang}, {Bai}, {Benisty}, {Birnstiel}, {Carpenter}, {Hughes}, {Ricci}, {Weaver}, \& {Wilner}}]{Huang_2018}
{Huang}, J., {Andrews}, S.~M., {Dullemond}, C.~P., {et~al.} 2018, \apjl, 869, L42, \dodoi{10.3847/2041-8213/aaf740}

\bibitem[{{K{\'o}sp{\'a}l} {et~al.}(2023){K{\'o}sp{\'a}l}, {{\'A}brah{\'a}m}, {Diehl}, {Banzatti}, {Bouwman}, {Chen}, {Cruz-S{\'a}enz de Miera}, {Green}, {Henning}, \& {Rab}}]{kospal_2023}
{K{\'o}sp{\'a}l}, {\'A}., {{\'A}brah{\'a}m}, P., {Diehl}, L., {et~al.} 2023, \apjl, 945, L7, \dodoi{10.3847/2041-8213/acb58a}

\bibitem[{{Law} {et~al.}(2021){Law}, {Loomis}, {Teague}, {{\"O}berg}, {Czekala}, {Andrews}, {Huang}, {Aikawa}, {Alarc{\'o}n}, {Bae}, {Bergin}, {Bergner}, {Boehler}, {Booth}, {Bosman}, {Calahan}, {Cataldi}, {Cleeves}, {Furuya}, {Guzm{\'a}n}, {Ilee}, {Le Gal}, {Liu}, {Long}, {M{\'e}nard}, {Nomura}, {Qi}, {Schwarz}, {Sierra}, {Tsukagoshi}, {Yamato}, {van't Hoff}, {Walsh}, {Wilner}, \& {Zhang}}]{Law_2021}
{Law}, C.~J., {Loomis}, R.~A., {Teague}, R., {et~al.} 2021, \apjs, 257, 3, \dodoi{10.3847/1538-4365/ac1434}

\bibitem[{{Long} {et~al.}(2019){Long}, {Herczeg}, {Harsono}, {Pinilla}, {Tazzari}, {Manara}, {Pascucci}, {Cabrit}, {Nisini}, {Johnstone}, {Edwards}, {Salyk}, {Menard}, {Lodato}, {Boehler}, {Mace}, {Liu}, {Mulders}, {Hendler}, {Ragusa}, {Fischer}, {Banzatti}, {Rigliaco}, {van de Plas}, {Dipierro}, {Gully-Santiago}, \& {Lopez-Valdivia}}]{Long_2019}
{Long}, F., {Herczeg}, G.~J., {Harsono}, D., {et~al.} 2019, \apj, 882, 49, \dodoi{10.3847/1538-4357/ab2d2d}

\bibitem[{{Mandell} {et~al.}(2012){Mandell}, {Bast}, {van Dishoeck}, {Blake}, {Salyk}, {Mumma}, \& {Villanueva}}]{Mandell_2012}
{Mandell}, A.~M., {Bast}, J., {van Dishoeck}, E.~F., {et~al.} 2012, \apj, 747, 92, \dodoi{10.1088/0004-637X/747/2/92}

\bibitem[{{Meijerink} {et~al.}(2009){Meijerink}, {Pontoppidan}, {Blake}, {Poelman}, \& {Dullemond}}]{Meijerink_2009}
{Meijerink}, R., {Pontoppidan}, K.~M., {Blake}, G.~A., {Poelman}, D.~R., \& {Dullemond}, C.~P. 2009, \apj, 704, 1471, \dodoi{10.1088/0004-637X/704/2/1471}

\bibitem[{Mirza-Romero {et~al.}(2023)Mirza-Romero, Banzatti, \& Öberg}]{MirzaRomero_2023}
Mirza-Romero, C.~E., Banzatti, A., \& Öberg, K.~I. 2023, iris (InfraRed Isothermal Slabs), 0.1.0,  Zenodo, \dodoi{10.5281/zenodo.10369000}

\bibitem[{{Najita} {et~al.}(2011){Najita}, {{\'A}d{\'a}mkovics}, \& {Glassgold}}]{Najita_2011}
{Najita}, J.~R., {{\'A}d{\'a}mkovics}, M., \& {Glassgold}, A.~E. 2011, \apj, 743, 147, \dodoi{10.1088/0004-637X/743/2/147}

\bibitem[{{Najita} {et~al.}(2018){Najita}, {Carr}, {Salyk}, {Lacy}, {Richter}, \& {DeWitt}}]{Najita_2018}
{Najita}, J.~R., {Carr}, J.~S., {Salyk}, C., {et~al.} 2018, \apj, 862, 122, \dodoi{10.3847/1538-4357/aaca39}

\bibitem[{{Najita} {et~al.}(2010){Najita}, {Carr}, {Strom}, {Watson}, {Pascucci}, {Hollenbach}, {Gorti}, \& {Keller}}]{Najita_2010}
{Najita}, J.~R., {Carr}, J.~S., {Strom}, S.~E., {et~al.} 2010, \apj, 712, 274, \dodoi{10.1088/0004-637X/712/1/274}

\bibitem[{{{\"O}berg} \& {Bergin}(2021)}]{Oberg_2021}
{{\"O}berg}, K.~I., \& {Bergin}, E.~A. 2021, \physrep, 893, 1, \dodoi{10.1016/j.physrep.2020.09.004}

\bibitem[{{{\"O}berg} {et~al.}(2011){{\"O}berg}, {Qi}, {Fogel}, {Bergin}, {Andrews}, {Espaillat}, {Wilner}, {Pascucci}, \& {Kastner}}]{oberg_2011}
{{\"O}berg}, K.~I., {Qi}, C., {Fogel}, J. K.~J., {et~al.} 2011, \apj, 734, 98, \dodoi{10.1088/0004-637X/734/2/98}

\bibitem[{{{\"O}berg} {et~al.}(2021){{\"O}berg}, {Guzm{\'a}n}, {Walsh}, {Aikawa}, {Bergin}, {Law}, {Loomis}, {Alarc{\'o}n}, {Andrews}, {Bae}, {Bergner}, {Boehler}, {Booth}, {Bosman}, {Calahan}, {Cataldi}, {Cleeves}, {Czekala}, {Furuya}, {Huang}, {Ilee}, {Kurtovic}, {Le Gal}, {Liu}, {Long}, {M{\'e}nard}, {Nomura}, {P{\'e}rez}, {Qi}, {Schwarz}, {Sierra}, {Teague}, {Tsukagoshi}, {Yamato}, {van't Hoff}, {Waggoner}, {Wilner}, \& {Zhang}}]{Oberg_2021b}
{{\"O}berg}, K.~I., {Guzm{\'a}n}, V.~V., {Walsh}, C., {et~al.} 2021, \apjs, 257, 1, \dodoi{10.3847/1538-4365/ac1432}

\bibitem[{{Pontoppidan} \& {Blevins}(2014)}]{Pontoppidan_2014}
{Pontoppidan}, K.~M., \& {Blevins}, S.~M. 2014, Faraday Discussions, 168, 49, \dodoi{10.1039/C3FD00141E}

\bibitem[{{Pontoppidan} {et~al.}(2010{\natexlab{a}}){Pontoppidan}, {Salyk}, {Blake}, \& {K{\"a}ufl}}]{Pontoppidan_2010b}
{Pontoppidan}, K.~M., {Salyk}, C., {Blake}, G.~A., \& {K{\"a}ufl}, H.~U. 2010{\natexlab{a}}, \apjl, 722, L173, \dodoi{10.1088/2041-8205/722/2/L173}

\bibitem[{{Pontoppidan} {et~al.}(2010{\natexlab{b}}){Pontoppidan}, {Salyk}, {Blake}, {Meijerink}, {Carr}, \& {Najita}}]{Pontoppidan_2010}
{Pontoppidan}, K.~M., {Salyk}, C., {Blake}, G.~A., {et~al.} 2010{\natexlab{b}}, \apj, 720, 887, \dodoi{10.1088/0004-637X/720/1/887}

\bibitem[{{Pontoppidan} {et~al.}(2023){Pontoppidan}, {Salyk}, {Banzatti}, {Zhang}, {Pascucci}, {Oberg}, {Long}, {Mirza-Romero}, {Carr}, {Najita}, {Blake}, {Arulanantham}, {Andrews}, {Ballering}, {Bergin}, {Calahan}, {Cobb}, {Colmenares}, {Dickson-Vandervelde}, {Dignan}, {Green}, {Heretz}, {Herczeg}, {Kalyaan}, {Krijt}, {Pauly}, {Pinilla}, {Trapman}, \& {Xie}}]{Pontoppidan_2023}
{Pontoppidan}, K.~M., {Salyk}, C., {Banzatti}, A., {et~al.} 2023, arXiv e-prints, arXiv:2311.17020, \dodoi{10.48550/arXiv.2311.17020}

\bibitem[{{Rieke} {et~al.}(2015){Rieke}, {Wright}, {B{\"o}ker}, {Bouwman}, {Colina}, {Glasse}, {Gordon}, {Greene}, {G{\"u}del}, {Henning}, {Justtanont}, {Lagage}, {Meixner}, {N{\o}rgaard-Nielsen}, {Ray}, {Ressler}, {van Dishoeck}, \& {Waelkens}}]{Rieke_2015}
{Rieke}, G.~H., {Wright}, G.~S., {B{\"o}ker}, T., {et~al.} 2015, \pasp, 127, 584, \dodoi{10.1086/682252}

\bibitem[{{Robitaille} {et~al.}(2021){Robitaille}, {Tollerud}, {Aldcroft}, {Bray}, {Van Kerkwijk}, {Droettboom}, {Sip{\H{o}}cz}, {Lim}, {Price-Whelan}, {Conseil}, {Dencheva}, {Bradley}, {Ginsburg}, {Mumford}, {Seifert}, {Craig}, {Mdmueller}, {StuartLittlefair}, {D'Avella}, {Lglattly}, {G{\"u}nther}, {Homeier}, {Donath}, {Perrygreenfield}, {Deil}, {Vin{\'\i}cius}, {Woillez}, {Vanderplas}, {Patil}, \& {N{\"o}the}}]{Robitaille_2021}
{Robitaille}, T., {Tollerud}, E., {Aldcroft}, T., {et~al.} 2021, {astropy/astropy: v4.2.1}, v4.2.1, Zenodo,  Zenodo, \dodoi{10.5281/zenodo.4670729}

\bibitem[{{Salyk} {et~al.}(2019){Salyk}, {Lacy}, {Richter}, {Zhang}, {Pontoppidan}, {Carr}, {Najita}, \& {Blake}}]{Salyk_2019}
{Salyk}, C., {Lacy}, J., {Richter}, M., {et~al.} 2019, \apj, 874, 24, \dodoi{10.3847/1538-4357/ab05c3}

\bibitem[{{Salyk} {et~al.}(2008){Salyk}, {Pontoppidan}, {Blake}, {Lahuis}, {van Dishoeck}, \& {Evans}}]{Salyk_2008}
{Salyk}, C., {Pontoppidan}, K.~M., {Blake}, G.~A., {et~al.} 2008, \apjl, 676, L49, \dodoi{10.1086/586894}

\bibitem[{{Salyk} {et~al.}(2011){Salyk}, {Pontoppidan}, {Blake}, {Najita}, \& {Carr}}]{Salyk_2011}
{Salyk}, C., {Pontoppidan}, K.~M., {Blake}, G.~A., {Najita}, J.~R., \& {Carr}, J.~S. 2011, \apj, 731, 130, \dodoi{10.1088/0004-637X/731/2/130}

\bibitem[{{Schwarz} {et~al.}(2023){Schwarz}, {Henning}, {Christiaens}, {Gasman}, {Samland}, {Perotti}, {Jang}, {Grant}, {Tabone}, {Morales-Calderon}, {Kamp}, {van Dishoeck}, {Gudel}, {Lagage}, {Argyriou}, {Barrado}, {Garatti}, {Glauser}, {Ray}, {Vandenbussche}, {Waters}, {Arabhavi}, {Kanwar}, {Olofsson}, {Rodgers-Lee}, {Schreiber}, \& {Temmink}}]{Schwarz_2023}
{Schwarz}, K.~R., {Henning}, T., {Christiaens}, V., {et~al.} 2023, arXiv e-prints, arXiv:2312.07135, \dodoi{10.48550/arXiv.2312.07135}

\bibitem[{{Sierra} {et~al.}(2021){Sierra}, {P{\'e}rez}, {Zhang}, {Law}, {Guzm{\'a}n}, {Qi}, {Bosman}, {{\"O}berg}, {Andrews}, {Long}, {Teague}, {Booth}, {Walsh}, {Wilner}, {M{\'e}nard}, {Cataldi}, {Czekala}, {Bae}, {Huang}, {Bergner}, {Ilee}, {Benisty}, {Le Gal}, {Loomis}, {Tsukagoshi}, {Liu}, {Yamato}, \& {Aikawa}}]{Sierra_2021}
{Sierra}, A., {P{\'e}rez}, L.~M., {Zhang}, K., {et~al.} 2021, \apjs, 257, 14, \dodoi{10.3847/1538-4365/ac1431}

\bibitem[{{Speagle}(2020)}]{Speagle_2020}
{Speagle}, J.~S. 2020, \mnras, 493, 3132, \dodoi{10.1093/mnras/staa278}

\bibitem[{{Stammler} {et~al.}(2023){Stammler}, {Lichtenberg}, {Drazkowska}, \& {Birnstiel}}]{Stammler_2023}
{Stammler}, S.~M., {Lichtenberg}, T., {Drazkowska}, J., \& {Birnstiel}, T. 2023, \aap, 670, L5, \dodoi{10.1051/0004-6361/202245512}

\bibitem[{{Stevenson} \& {Lunine}(1988)}]{Stevenson_1988}
{Stevenson}, D.~J., \& {Lunine}, J.~I. 1988, \icarus, 75, 146, \dodoi{10.1016/0019-1035(88)90133-9}

\bibitem[{{Tabone} {et~al.}(2023){Tabone}, {Bettoni}, {van Dishoeck}, {Arabhavi}, {Grant}, {Gasman}, {Henning}, {Kamp}, {G{\"u}del}, {Lagage}, {Ray}, {Vandenbussche}, {Abergel}, {Absil}, {Argyriou}, {Barrado}, {Boccaletti}, {Bouwman}, {Caratti o Garatti}, {Geers}, {Glauser}, {Justannont}, {Lahuis}, {Mueller}, {Nehm{\'e}}, {Olofsson}, {Pantin}, {Scheithauer}, {Waelkens}, {Waters}, {Black}, {Christiaens}, {Guadarrama}, {Morales-Calder{\'o}n}, {Jang}, {Kanwar}, {Pawellek}, {Perotti}, {Perrin}, {Rodgers-Lee}, {Samland}, {Schreiber}, {Schwarz}, {Colina}, {{\"O}stlin}, \& {Wright}}]{Tabone_2023}
{Tabone}, B., {Bettoni}, G., {van Dishoeck}, E.~F., {et~al.} 2023, Nature Astronomy, 7, 805, \dodoi{10.1038/s41550-023-01965-3}

\bibitem[{{van Dishoeck} {et~al.}(2014){van Dishoeck}, {Bergin}, {Lis}, \& {Lunine}}]{van_Dishoeck_2014}
{van Dishoeck}, E.~F., {Bergin}, E.~A., {Lis}, D.~C., \& {Lunine}, J.~I. 2014, in Protostars and Planets VI, ed. H.~{Beuther}, R.~S. {Klessen}, C.~P. {Dullemond}, \& T.~{Henning}, 835--858, \dodoi{10.2458/azu_uapress_9780816531240-ch036}

\bibitem[{{Wells} {et~al.}(2015){Wells}, {Pel}, {Glasse}, {Wright}, {Aitink-Kroes}, {Azzollini}, {Beard}, {Brandl}, {Gallie}, {Geers}, {Glauser}, {Hastings}, {Henning}, {Jager}, {Justtanont}, {Kruizinga}, {Lahuis}, {Lee}, {Martinez-Delgado}, {Mart{\'\i}nez-Galarza}, {Meijers}, {Morrison}, {M{\"u}ller}, {Nakos}, {O'Sullivan}, {Oudenhuysen}, {Parr-Burman}, {Pauwels}, {Rohloff}, {Schmalzl}, {Sykes}, {Thelen}, {van Dishoeck}, {Vandenbussche}, {Venema}, {Visser}, {Waters}, \& {Wright}}]{Wells_2015}
{Wells}, M., {Pel}, J.~W., {Glasse}, A., {et~al.} 2015, \pasp, 127, 646, \dodoi{10.1086/682281}

\bibitem[{{Woitke} {et~al.}(2019){Woitke}, {Kamp}, {Antonellini}, {Anthonioz}, {Baldovin-Saveedra}, {Carmona}, {Dionatos}, {Dominik}, {Greaves}, {G{\"u}del}, {Ilee}, {Liebhardt}, {Menard}, {Min}, {Pinte}, {Rab}, {Rigon}, {Thi}, {Thureau}, \& {Waters}}]{Woitke_2019}
{Woitke}, P., {Kamp}, I., {Antonellini}, S., {et~al.} 2019, \pasp, 131, 064301, \dodoi{10.1088/1538-3873/aaf4e5}

\bibitem[{{Zhang} {et~al.}(2013){Zhang}, {Pontoppidan}, {Salyk}, \& {Blake}}]{Zhang_2013}
{Zhang}, K., {Pontoppidan}, K.~M., {Salyk}, C., \& {Blake}, G.~A. 2013, \apj, 766, 82, \dodoi{10.1088/0004-637X/766/2/82}

\end{thebibliography}
\bibliographystyle{aasjournal}



\end{document}